\documentclass[journal=langd5,manuscript=article,layout=twocolumn]{achemso}

\pdfoutput=1

\let\oldmaketitle\maketitle
\let\maketitle\relax

\makeatletter
\let\l@addto@macro\relax
\makeatother
\usepackage[fontsize=10pt]{scrextend}

\usepackage{amsmath}
\usepackage{amssymb}
\usepackage{color}
\usepackage{graphicx}
\usepackage{enumerate}
\usepackage{epstopdf}
\usepackage{mathtools}
\usepackage{texdraw}
\usepackage{tikz}
\usepackage{transparent}
\usepackage{textcomp}
\usepackage{gensymb}
\usepackage{graphicx}
\usepackage[separate-uncertainty=true]{siunitx}
\usepackage{placeins}
\usepackage{lipsum}
\usepackage{nameref}
\usepackage[hidelinks]{hyperref}
\usepackage[export]{adjustbox}

\renewcommand{\d}[1]{\mathrm{d}#1}
\newcommand{\dd}[2]{\frac{\mathrm{d}#1}{\mathrm{d}#2}}

\newcommand{\taurlx}{\tau_\mathrm{rlx}}
\newcommand{\tevap}{t_\mathrm{evap}}
\newcommand{\tauevap}{\tau_\mathrm{evap}}
\newcommand{\ratio}{\tauevap/\taurlx}
\newcommand{\gammaLG}{\gamma_\mathrm{LG}}
\newcommand{\gammaSL}{\gamma_\mathrm{SL}}
\newcommand{\gammaSG}{\gamma_\mathrm{SG}}
\newcommand{\thetaeq}{\theta_\mathrm{eq}}
\newcommand{\thetar}{\theta_\mathrm{r}}
\newcommand{\thetaa}{\theta_\mathrm{a}}
\newcommand{\fc}{f_\mathrm{c}}
\newcommand{\fp}{f_\mathrm{p}}

\graphicspath{{Figures/}}
\author{Thijs W.G. van der Heijden}
\email{t.w.g.van.der.heijden@tue.nl}
\affiliation{Department of Applied Physics, Eindhoven University of Technology, P.O. Box 513,\\5600 MB Eindhoven, The Netherlands}

\author{Anton A. Darhuber}
\affiliation{Department of Applied Physics, Eindhoven University of Technology, P.O. Box 513,\\5600 MB Eindhoven, The Netherlands}

\author{Paul van der Schoot}
\affiliation{Department of Applied Physics, Eindhoven University of Technology, P.O. Box 513,\\5600 MB Eindhoven, The Netherlands}
\alsoaffiliation{Instituut voor Theoretische Fysica, Universiteit Utrecht, Princetonplein 5,\\3584 CC Utrecht, The Netherlands}

\title{A macroscopic model for sessile droplet evaporation\\on a flat surface}

\begin{document}
\newlength{\figwidth}
\setlength{\figwidth}{\linewidth}

\twocolumn[
\begin{@twocolumnfalse}
\oldmaketitle
\begin{abstract}
\noindent The evaporation of sessile droplets on a flat surface involves a complex interplay between phase change, diffusion, advection and surface forces. In an attempt to significantly reduce the complexity of the problem and to make it manageable, we propose a simple model hinged on a surface free energy-based relaxation dynamics of the droplet shape, a diffusive evaporation model and a contact line pinning mechanism governed by a yield stress. Our model reproduces the known dynamics of droplet shape relaxation and of droplet evaporation, both in the absence and in the presence of contact line pinning. We show that shape relaxation during evaporation significantly affects the lifetime of a drop. We find that the dependence of the evaporation time on the initial contact angle is a function of the competition between the shape relaxation and evaporation, and is strongly affected by any contact line pinning.
\vspace{10pt}
\end{abstract}
\end{@twocolumnfalse}
]

\section{Introduction}
\label{sec:introduction}
Understanding the dynamics of spreading and drying of droplets deposited on a substrate is of importance to many practices, 
such as inkjet printing~\cite{Kuang2014,Park2006,Calvert2001}, pesticide spraying~\cite{Yu2009} and semiconductor device manifacturing~\cite{Wei2009,Belmiloud2012}. In the semiconductor industry, photolithographic methods are employed to define patterns for integrated circuits on wafers, coated with photosensitive polymer layers~\cite{Mack2007,Wei2009}. Often, water immersion is used to increase the resolution of the lithography process~\cite{Lin1987,Mack2007}. However, if any droplets are left behind on a wafer, they may induce so-called watermark defects in the photoresist layer~\cite{Wei2009,Kocsis2006}.

Due to the importance of understanding drying processes, the drying of droplets on surfaces has been intensely studied experimentally~\cite{Bourges-Monnier1995,Uno1998,Cachile2002,Fukai2006,Belmiloud2012,Yu2012,Hu2002}, theoretically~\cite{Picknett1977,Erbil2002,Hughes2015,Hu2002,Man2016} and numerically~\cite{Frank2012,Ledesma-Aguilar2014,Hirvi2006,Zhang2014}. Nevertheless, the understanding of this multifaceted problem remains incomplete due to the multitude of coupled processes that determine the evaporation dynamics. Apart from the evaporation itself, processes such as convection and heat transport in the droplet, shape relaxation, and contact line pinning play a role. 

Associated with the complex physics of the problem at hand are a large number of physical parameters, the relative importance of which depends on the initial and boundary conditions as well as the time and length scales of interest. Therefore, we aim to develop a macroscopic model that does not resolve the details of, \latin{e.g.}, the velocity field inside the droplet or the vapour concentration field around it. Rather, we consider three constituents to make up our model: (1) interfacial free energy-based relaxation for the droplet shape, (2) diffusion-limited evaporation and (3) contact line pinning.

In the literature, various authors studied the evaporation of droplets, focusing on two limiting modes of evaporation: a droplet evaporates with either a constant contact area or a constant contact angle, allowing transitions between these limits~\cite{Picknett1977,Fukai2006,Stauber2014,Stauber2015a}. Others have investigated the shape relaxation of droplets by measuring the contact angle of non-evaporating droplets in time~\cite{Schonhorn1966,Newman1968,Blake1969}. Our model combines both aspects, so it is not restricted to the two evaporation modes but also captures the shape relaxation of the droplet during the evaporation process. It extends the evaluation of \citet{Stauber2014} by taking into account the contact line dynamics, \latin{i.e.}, incorporating both advancing and receding contact lines and considering cases without contact line pinning. Moreover, to describe the transition between mobile and pinned contact lines, our model includes a yield stress that governs contact line pinning: contact line motion is inhibited for capillary driving forces below a critical stress.

The remainder of this paper is organised as follows. In the \nameref{sec:theory} section, we present the main ingredients of our phenomenological model. The \nameref{sec:results} section contains an overview of representative cases of evaporation with and without contact line pinning, as well as a comparison with experimental data. In the \nameref{sec:discussion} section we discuss in detail the implications of choices made for certain parameters during the calculations. In the \nameref{sec:conclusion} section, we summarise our results and present our main conclusions.

\section{Theory}
\label{sec:theory}
\begin{figure}[htp!]
\centering
\large
\def\svgwidth{\figwidth}
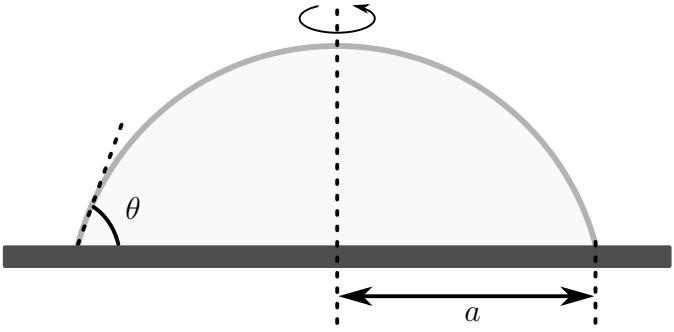
\caption{Schematic of an axisymmetric droplet on a planar surface, with contact angle ${\theta}$ and radius ${a}$ of the contact area.}
\label{fig:droplet_schematic}
\end{figure}

The focus of this work is on droplets of sizes smaller than the capillary length ${l_c=\sqrt{\gammaLG/\rho g}}$, which allows us to model the droplet as a spherical cap~\cite{DeGennes2004}. Here, ${\gammaLG}$ denotes the surface tension of the liquid-gas interface, ${\rho}$ the mass density of the fluid and ${g}$ the gravitational acceleration. For water in air at room temperature, ${l_c\simeq \SI{3}{\milli\metre}}$~\cite{DeGennes2004}. We presume the liquid to be incompressible. If the shape of the droplet is described as a spherical cap, it is uniquely defined by only two parameters. We choose for these the radius ${a}$ of the contact area and the contact angle ${\theta}$ of the drop with the solid surface that we assume to be rigid (see Fig.~\ref{fig:droplet_schematic})~\cite{Extrand1996}. Geometrically, they are related to the droplet volume ${V}$ according to
\begin{equation}
\label{eq:volume}
V(a,\theta)=\pi a^3 \left(\frac{2-3\cos\theta+\cos^3\theta}{3\sin^3\theta}\right).
\end{equation}

Equation~\eqref{eq:volume} implies that, for a given volume ${V}$, a prescribed value for the contact area radius ${a}$ defines the contact angle ${\theta}$, and \latin{vice versa}. Within a macroscopic description of the droplet shape, the equilibrium values of ${a}$ and ${\theta}$ are determined by $\gammaLG$, as well as the the solid-liquid and solid-gas interfacial tensions, ${\gammaSL}$ and ${\gammaSG}$. We associate the droplet shape with an interfacial free energy ${{F}(a,\theta)}$, given by the sum of the interfacial tensions multiplied by the respective surface areas,
\begin{equation}
\label{eq:free_energy}
{F}(a,\theta)=\pi a^2 \left(\gammaSL-\gammaSG+\frac{2\gammaLG}{1+\cos\theta}\right).
\end{equation}
For a given volume ${V}$, Eq.~\eqref{eq:free_energy} can be expressed as a function of ${\theta}$ only, using Eq.~\eqref{eq:volume}, that is, ${{F}(a,\theta)\to F(\theta)}$~\cite{Whyman2008}. Minimising this free energy $F$ with respect to ${\theta}$ produces the following equation for its optimal value,
\begin{equation}
\gammaSG-\gammaSL-\gammaLG\cos\thetaeq=0,
\label{eq:young}
\end{equation} 
which is the well-known Young's equation for the equilibrium contact angle ${\thetaeq}$~\cite{Young1805,DeGennes1985}.

Out of equilibrium, Eq.~\eqref{eq:young} does not hold. To describe how an out-of-equilibrium droplet shape relaxes towards equilibrium, we construct a kinetic equation using a relaxational dynamics approach based on our free energy landscape~\cite{Swift1977,Goldenfeld1992}. This allows us to quite naturally include the effects of steady evaporation and of a potential pinning of the contact line. In the next subsections we discuss separately and in detail the three main components of our phenomenological model: the relaxation dynamics of the droplet shape, the description for diffusive evaporation and the contact line pinning mechanism.

\subsection{Relaxation dynamics of the droplet shape}
\label{subsec:theory_relax}
Based on the free energy ${F(\theta)}$ obtained from Eqs.~(\ref{eq:volume},\ref{eq:free_energy}), we construct a relaxation equation for the contact angle ${\theta}$,
\begin{equation}
\label{eq:dyn_theta}
\dd{\theta}{t}=-\Gamma\dd{F}{\theta},
\end{equation} 
where ${\Gamma}$ is a phenomenological relaxation rate that we specify in more detail below. Equation~\eqref{eq:dyn_theta} describes a rate of change that is proportional to the generalised force ${\d{F}/\d{\theta}}$. This is in analogy to so-called model A dynamics commonly applied in the kinetics of phase transitions of non-conserved order parameters~\cite{Swift1977,Goldenfeld1992,Chaikin1995}. We note that it is also possible to derive kinetic equations by equating capillary and viscous forces~\cite{Harth2012,Weirich2017}. In our model, this balance is implicit in the parameter ${\Gamma}$. 

Several experimental and theoretical works have identified the difference between the cosines of the instantaneous, time-dependent contact angle ${\theta}$ and its equilibrium value ${\thetaeq}$, given by Eq.~\eqref{eq:young}, to be the driving force for the motion of the contact line~\cite{Newman1968,Blake1969,DeGennes1985,Bonn2009,Andrieu2002}. For small values of the difference ${\cos\theta-\cos\thetaeq}$, this relaxation can be described by a simple exponential function~\cite{Newman1968,Cherry1969}. The exponential decay allows us to identify a characteristic timescale ${\taurlx}$ as
\begin{equation}
\label{eq:exp_relax}
\cos\theta-\cos\thetaeq\propto \exp(-t/\taurlx).
\end{equation}
We discuss the functional expression for ${\taurlx}$ below.

To make contact with Eq.~\eqref{eq:exp_relax}, we transform Eq.~\eqref{eq:dyn_theta} into a kinetic equation for $\cos\theta$, and expand it around the equilibrium $\cos\thetaeq$. A linearisation produces an exponentially decaying ${\cos\theta}$, from which we determine ${\Gamma}$. This yields
\begin{equation}
\label{eq:Gamma}
\Gamma=\frac{1}{\taurlx\gammaLG V^{2/3} \Theta(\cos\thetaeq)\alpha(t)},
\end{equation}
with 
\begin{equation}
\Theta(x)\equiv 2(9\pi)^{1/3}\left(1-x\right)^3\left(1+x\right)\left(2-3x+x^3\right)^{-5/3}.
\end{equation}
Equations~(\ref{eq:dyn_theta},\ref{eq:Gamma}) reproduce Eq.~\eqref{eq:exp_relax} for small deviations from equilibrium. In Eq.~\eqref{eq:Gamma}, we introduce a time-dependent, dimensionless factor $\alpha(t)$ to account for changes in the dynamics of the droplet shape relaxation due to, \latin{e.g.}, a change in droplet size over time. We return to this below. Equations~(\ref{eq:dyn_theta},\ref{eq:Gamma}) describe how a droplet deposited with a non-equilibrium initial contact angle ${\theta_0}$ relaxes to the equilibrium value ${\thetaeq}$ in a relaxation process characterised by a fundamental timescale ${\taurlx}$.

The characteristic timescale $\taurlx$ for the shape relaxation has been identified in various experimental and theoretical works to be dependent on the fluid viscosity $\eta$, the liquid-gas interfacial tension $\gammaLG$ and a length scale $L$~\cite{Schonhorn1966,Newman1968,Cherry1969,VanOene1969,Varma1974,Andrieu2002,Weirich2017,Man2016}, as
\begin{equation}
\label{eq:taurlx}
\taurlx = \frac{\eta L}{\gammaLG}.
\end{equation}
In experiments on spreading of polymeric fluids, this length scale $L$ has been described as a measure for the slip or friction length of the interaction between a polymeric liquid and the solid~\cite{Schonhorn1966,Newman1968,Cherry1969}, which seems to be independent of droplet dimensions~\cite{Newman1968} and has been estimated to be of the order of micrometres~\cite{Cherry1969}. In works on the coalescence of droplets, however, the length scale $L$ has been shown to be proportional to the droplet size $R$~\cite{Andrieu2002,Weirich2017,Man2016}, which seems in agreement with more recent experimental and theoretical work on the spreading of polymer melts~\cite{VanOene1969} and spherical droplets of simple liquids~\cite{Varma1974}. For this reason, we take the length scale $L$ to be $k{V_0^{1/3}}$, with $V_0$ the initial volume of the droplet, making it proportional to the droplet size, and $k$ a dimensionless proportionality constant that can be related to an Arrhenius factor~\cite{Andrieu2002}. Hence,
\begin{equation}
\label{eq:tau_rel_1}
\taurlx=k\frac{\eta V_0^{1/3}}{\gammaLG}.
\end{equation}

As the droplet size decreases during evaporation, the length scale $L$ related to the shape relaxation may (1) remain constant (in the case that $L$ is related to a slip or friction length), or (2) decrease with the droplet size. The scale factor $\alpha(t)$ can be employed to incorporate either behaviour into the dynamics described by Eqs.~(\ref{eq:dyn_theta},\ref{eq:Gamma}). If $L$ remains constant, we may choose $\alpha=1$, whereas for a size-dependent length scale ${\alpha(t)=\left(V(t)/V_0\right)^{1/3}}$. As we shall see, it turns out that the two expressions for $\alpha$ do give rise to small differences in the droplet dynamics albeit that the lifetime of an evaporating droplet is not sensitive to whether $\alpha$ is proportional to the droplet size or not. For simplicity, we set $\alpha=1$ for the evaluation of our results. We discuss the implications of choosing the alternative $\alpha(t)$ in more detail in the \nameref{sec:discussion} section.

The structure of Eq.~\eqref{eq:Gamma} allows for the implementation of different models for droplet shape relaxation, as long as it progresses exponentially in the limit of small deviations from equilibrium, as in Eq.~\eqref{eq:exp_relax}. For example, from a microscopic perspective, the motion of the contact line is often described by a so-called molecular kinetic theory (MKT)~\cite{Blake1969}. This theory describes the motion of the contact line in terms of small jumps over the intrinsically microscopically inhomogeneous surface, driven by thermal fluctuations. It has been shown to predict contact line dynamics in agreement with experiments and molecular simulation~\cite{Blake1969,DeRuijter1997,Blake2006,Seveno2009,Blake1997}. MKT relates the velocity $\d{a}/\d{t}$ of the contact line to the driving force via the expression
\begin{equation}
\label{eq:MKT}
\dd{a}{t}=\frac{2\xi k_\mathrm{B} T \exp\left(-\frac{G^*}{k_\mathrm{B} T}\right)}{\eta v_\mathrm{L}} \sinh\left[\frac{\gammaLG\xi^2}{2k_\mathrm{B} T}\left(\cos\thetaeq-\cos\theta\right)\right],
\end{equation}
where ${\xi}$ denotes the distance between adsorption sites on the surface, ${k_\mathrm{B} T}$ the usual thermal energy with ${k_\mathrm{B}}$ Boltzmann's constant and ${T}$ the absolute temperature, ${v_\mathrm{L}}$ the molecular volume of the liquid and ${G^*}$ the surface contribution to the activation free energy of wetting~\cite{Blake1969,Blake2002,Blake2006,Snoeijer2013}.

If we translate Eq.~\eqref{eq:MKT} in terms of the time evolution of the cosine of the contact angle, \latin{i.e.}, make use of Eq.~\eqref{eq:volume}, and expand this to linear order for small deformations ${\cos\theta-\cos\thetaeq}$, we find that the characteristic timescale ${\taurlx}$ according to molecular kinetic theory must be given by
\begin{equation}
\label{eq:tau_mkt}
\taurlx=\frac{\eta V^{1/3}}{\gammaLG}\frac{v_\mathrm{L} \left(\frac{3}{\pi}\frac{(1+\cos\thetaeq)\sin\thetaeq}{2-\cos\thetaeq-\cos^2\thetaeq}\right)^{1/3}}{\xi^3 \exp\left(-\frac{G^*}{k_\mathrm{B} T}\right)(2+\cos\thetaeq)\sin^2\thetaeq}.
\end{equation}

We see that the functional form of $\taurlx$ of Eq.~\eqref{eq:tau_mkt} is analogous to that of Eq.~\eqref{eq:tau_rel_1}. This suggests that the characteristic shape relaxation time as predicted by MKT, which is a microscopic theory in origin, to linear order also is a function of macroscopic parameters such as droplet size, viscosity and surface tension.

This concludes our analysis of the relaxation dynamics of small drops. We next describe how quasi-steady evaporation affects the dynamics of a deposited droplet, presuming that an instantaneous free energy can be defined, in effect presuming a separation of timescales. 

\subsection{Evaporation of the droplet}
\label{subsec:theory_evap}
We take quasi-stationary, isothermal vapour diffusion into the surrounding gas phase to be the governing mechanism for evaporation, assuming the droplet to be in contact with an infinite volume of gas. \citet{Picknett1977} derived an expression for the rate of change in mass of a droplet as a function of the contact angle ${\theta}$. The rate of change of the volume ${V}$ of a droplet can then be written as
\begin{equation}
\label{eq:evap}
\dd{V}{t}=-\frac{2\pi a D \Delta c}{\rho \sin\theta} f(\theta),
\end{equation}
where ${D}$ denotes the diffusion coefficient of vapour molecules in the gas phase and ${\rho}$ the mass density of the liquid~\cite{Erbil2002}. Furthermore, ${\Delta c\equiv c_\mathrm{s} - c_\infty}$ denotes the difference between the vapour mass concentration ${c_s}$ near the liquid-gas interface (in units of mass per volume), presumed to be the saturation value of the fluid molecules in the gas phase, and the vapour mass concentration ${c_\infty}$ at infinity, \latin{i.e.}, that of the ambient atmosphere. Finally, ${f(\theta)}$ denotes a geometric factor for which an exact analytical expression is not available in closed form~\cite{Popov2005,Picknett1977}. For our purposes a polynomial representation for ${f(\theta)}$,
\begin{equation}
\label{eq:evap_f}
f(\theta) = 
\left\lbrace
\begin{matrix}
\begin{matrix}
0.6366\theta + 0.09591\theta^2 \\- 0.06144\theta^3 
\end{matrix} & \mathrm{for}\ 0\leq\theta<0.175,
\\ \\
\begin{matrix}
0.00008957 + 0.6333\theta \\+ 0.116\theta^2 - 0.08878\theta^3 \\+ 0.01033\theta^4
\end{matrix}
& \mathrm{for}\ 0.175\leq\theta\leq \pi.
\end{matrix}
\right.
\end{equation}
is sufficiently accurate. Indeed, the error of the approximant is less than ${0.2\%}$ for all values of ${\theta}$~\cite{Picknett1977}.

For a constant contact angle ${\theta}$, Eq.~\eqref{eq:evap} can be expressed entirely in the contact area radius ${a(t)}$ using Eq.~\eqref{eq:volume} and solved exactly. This gives 
\begin{equation}
\label{eq:a_t}
a(t) = \sqrt{a_0^2-\frac{4D\Delta c}{\rho} \frac{f(\theta)\sin^2\theta}{2-3\cos\theta+\cos^3\theta}\,t},
\end{equation}
where ${a_0}$ denotes the initial value of the contact area radius ${a_0=a(0)}$. It shows that the contact area ${\pi a^2}$ decreases linearly in time, a known experimental result~\cite{Erbil2002}. From Eq.~\eqref{eq:a_t}, we deduce that the time ${\tevap}$ it takes to evaporate a droplet is the longest for ${\theta=\pi/2}$. For this contact angle, the evaporation time ${\tauevap}$ is given by the simple expression
\begin{equation}
\label{eq:tau_evap}
\tauevap=\frac{\rho}{2D \Delta c}\left(\frac{3V_0}{2\pi}\right)^{2/3}.
\end{equation}
In the remainder of this work, we shall scale all evaporation times to ${\tauevap}$. Note that this point we have identified the two fundamental timescales that describe our problem: the fundamental relaxation time $\taurlx$, Eq.~\eqref{eq:tau_rel_1}, and the fundamental evaporation time $\tauevap$, Eq.~\eqref{eq:tau_evap}. The actual evaporation time depends on the one hand on the initial contact angle and the relaxation dynamics of the droplet shape, but also on whether or not contact line pinning takes place.

\subsection{Contact line pinning}
Contact line pinning is the phenomenon where the contact line of the droplet becomes stuck, permanently or temporarily, on structural or chemical inhomogeneities of the supporting surface~\cite{Johnson1964,Andersen1996,Brandon1996, Schaffer1998,Ondarcuhu2005,Bonn2009,Debuisson2016,DeConinck2017}. In general, a droplet in the pinned state exhibits a contact angle different from the equilibrium angle ${\thetaeq}$, as it cannot relax to its equilibrium shape. We model the influence of surface heterogeneities by introducing a net macroscopic threshold force per unit length, ${\fp}$, exerted in the plane of the surface along the radial direction of the circular contact line. It has a direction opposite to the capillary driving force per unit length, ${\fc}$. As both ${\fp}$ and ${\fc}$ are exerted on the perimeter of the contact area, we for simplicity refer to both as a force.

If the magnitude of the capillary driving force is smaller than the threshold ${\fp}$, then the contact line remains pinned. On the other hand, if it is greater, we allow the contact line to move. The resulting contact line motion is quasi-steady and hence the associated friction does not depend on the velocity of the contact line. For simplicity, we presume that the yield force ${\fp}$ does not depend on the position on the surface. The capillary force we define as
\begin{equation}
\label{eq:cl_force_density}
\fc =-\frac{1}{2\pi a}\dd{F}{a}=-\gammaLG\left(\cos\theta-\cos\thetaeq\right),
\end{equation}
where we have made use of Young's law, Eq.~\eqref{eq:young}. In our prescription, we allow motion of the contact line as long as ${\left|\fc\right| > \fp}$. Equation~\eqref{eq:cl_force_density} is also referred to as the unbalanced Young force or unbalanced capillary forces~\cite{DeGennes1985,Stauber2015a,Snoeijer2013}. 

The magnitude of the pinning force ${\fp}$ defines a contact angle range in which the capillary force ${\fc}$ is too weak to overcome pinning. As long as the contact angle $\theta$ resides within this range, the contact area remains constant. We refer to this range as the ``fixed area'' regime and it turns out to be bounded by the receding and advancing contact angles, ${\thetar}$ and ${\thetaa}$. Within our model, the values of these quantities depend on the pinning force ${\fp}$~\cite{Bonn2009}, according to
\begin{align}
\label{eq:theta_r}
\thetar&=\arccos\left(\cos\thetaeq+{\fp}/{\gammaLG}\right),\\
\label{eq:theta_a}
\thetaa&=\arccos\left(\cos\thetaeq-{\fp}/{\gammaLG}\right).
\end{align}
The receding and advancing contact angles indicate the points at which the pinning-depinning transitions occur. If the droplet evaporates while initially being in the pinned (fixed area) state, the contact angle decreases until the droplet depins at a value equal to ${\thetar}$, after which the evaporation continues with a constant contact angle $\thetar$ and a receding contact line. In contrast to a constant ${\thetar}$, a constant advancing angle ${\thetaa}$ is not encountered for droplets with decreasing volume, but it can only be observed as the point at which the droplet becomes pinned after initial spreading. 

\section{Results}
\label{sec:results}
We now compare predictions of our phenomenological model with the full non-linear response presented by molecular kinetic theory (MKT) and with experiments on droplet evaporation in the presence of contact line pinning. We quantify the competition between evaporation and relaxation using the ratio of the two timescales ${\tauevap}$ and ${\taurlx}$. It determines, together with the initial and equilibrium contact angles as well as the magnitude of the pinning force, the lifetime of an evaporating droplet. Both fundamental timescales depend only on the properties of the fluid and the surrounding vapour phase. The ratio of the two timescales has also been addressed by \citet{Man2016} to be important in the context of evaporation problems. The ratio ${\ratio}$ is analogous to the inverse of the parameter $k_\mathrm{ev}$ presented in Ref.~\citenum{Man2016}, which ranges in magnitude from less than $10^{-3}$ to more than $10^{-1}$. We choose the droplets to be hemispherical in equilibrium, \latin{i.e.}, ${\thetaeq=\pi/2}$, which is typical for a water droplet on a polymer substrate. The implications of choosing a different equilibrium contact angle are discussed in the \nameref{sec:discussion} section.

Because sessile droplet shape relaxation and evaporation have been described separately in the literature before, we feel it instructive to first investigate how our model compares to those works and to known experimental data. Our more complete model combines shape relaxation, droplet evaporation and contact line pinning, and we discuss its predictions subsequently.

\subsection{Shape relaxation and pinning-depinning transition}
\begin{figure}[htp!]
\centering
\includegraphics[width=\figwidth]{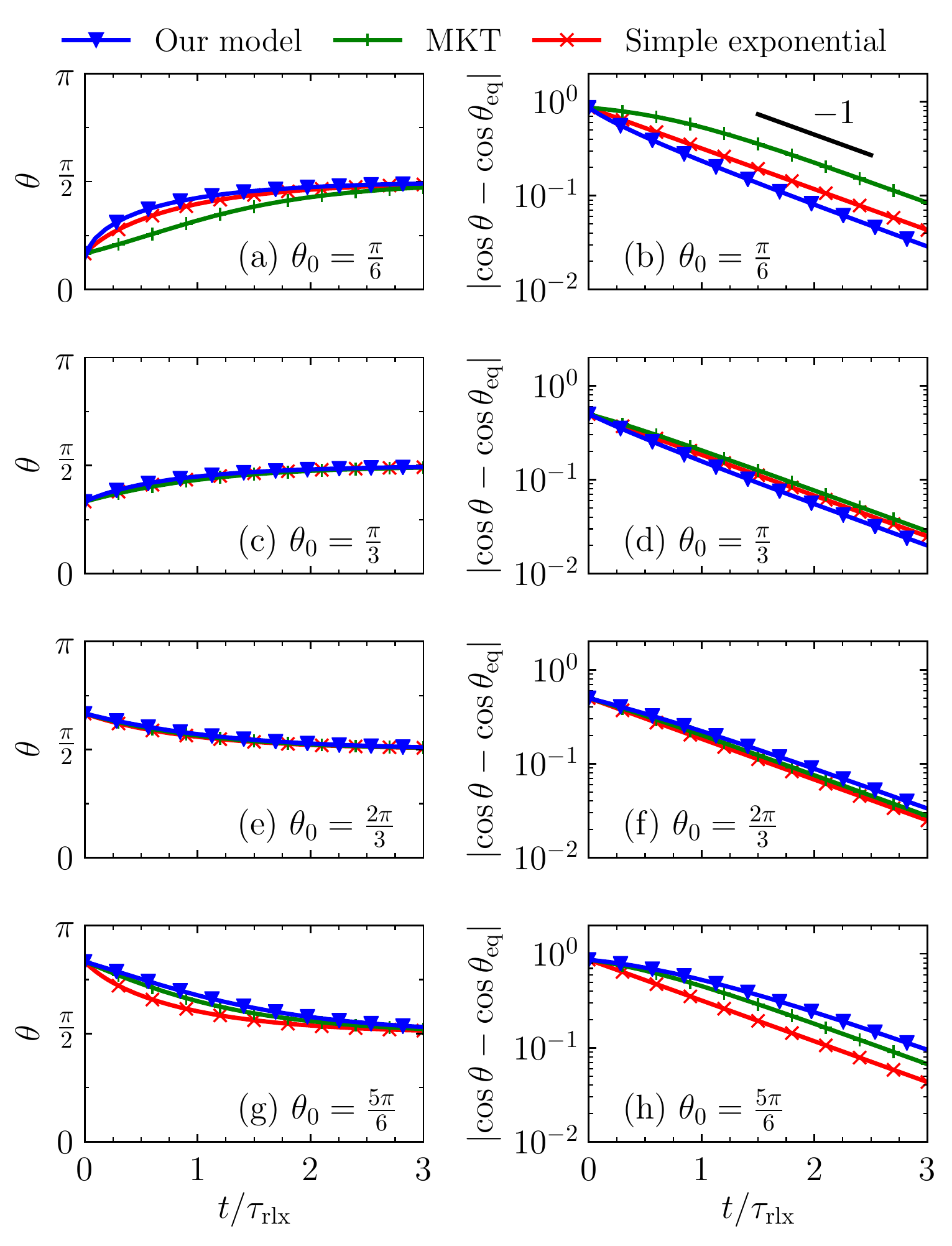}
\caption{Comparison between predictions for the relaxation dynamics of the contact angle ${\theta}$ to its equilibrium value ${\thetaeq=\pi/2}$ from our model (blue triangles), molecular kinetic theory (green pluses) and a simple exponential relaxation for comparison (red crosses). Left: contact angle ${\theta}$ as function of dimensionless time ${t/\taurlx}$, right: the corresponding ${\left|\cos\theta-\cos\thetaeq\right|}$.}
\label{fig:relax_MKT_exp}
\end{figure}

To illustrate the relaxation dynamics predicted by our free energy-based model and to compare the predictions to an existing model for contact line dynamics, we first compare our theory with the relaxation dynamics of a droplet deposited on a surface according to molecular kinetic theory (MKT). This theory, which has a microscopic basis, is shown to describe experimentally measured contact line dynamics rather well~\cite{Blake1969,DeRuijter1997,Blake2006,Seveno2009}. As discussed in the \nameref{sec:theory} section, for small values of ${\cos\theta-\cos\thetaeq}$, MKT predicts an exponential relaxation with a timescale $\taurlx$ given by Eq.~\eqref{eq:tau_mkt}. For greater values, however, the dynamics deviates from a simple single exponential description. In order to compare the non-linear contact angle dynamics predicted by our model to that described by MKT, we solve Eq.~\eqref{eq:MKT} numerically. For convenience, we set the equilibrium contact angle to a value of ${\thetaeq=\pi/2}$ and choose four initial angles ${\theta_0}$ symmetrically around this angle. In Fig.~\ref{fig:relax_MKT_exp}, we compare the time dependence of the contact angle ${\theta}$ and the absolute value of the difference between the cosines of $\theta(t)$ and $\thetaeq$. Indicated in the figures are the results of our model (blue triangles), the MKT result (green pluses) and a simple single exponential relaxation as given in Eq.~\eqref{eq:exp_relax} (red crosses). Note that, as in all cases ${\taurlx}$ and ${\thetaeq}$ are fixed, the results indicated in the figures as ``Our model'' are independent of the choice of the characteristic timescale and hence also describe the result for, \latin{e.g.}, the linearised version of the MKT model.

Figure~\ref{fig:relax_MKT_exp} teaches us the following:
\begin{enumerate}[i.]
\item For deviations of ${\pm \pi/6}$ from the equilibrium value of ${\pi/2}$, see Fig.~\ref{fig:relax_MKT_exp}(c) and (e),
agreement between the evolution of the contact angle as a function of scaled time predicted by all three descriptions is excellent. For greater initial deviations from the equilibrium angle, shown in Figures~\ref{fig:relax_MKT_exp}(a) and (g), agreement remains remarkably good, in particular for the larger initial angle;

\item Figures~\ref{fig:relax_MKT_exp}(d) and (f) highlight any inconsistencies for small deviations from the equilibrium by focussing on the difference of the cosines on a logarithmic scale. These figures show that well within one characteristic timescale simple single exponential decay is reached. Any small late-stage deviations between the curves is caused by the early-stage non-linear behaviour. Figures~\ref{fig:relax_MKT_exp}(b) and (h) show that even for greater initial deviations from the equilibrium contact angle, simple single exponential decay occurs within one characteristic timescale.
\end{enumerate}

The process of droplet evaporation in the presence of contact line pinning has been studied theoretically by \citet{Stauber2014}, who describe the dependence of the evaporation time ${\tevap}$ on the initial contact angle ${\theta_0}$, where they fix the receding contact angle ${\thetar}$ to several values. They consider two separate modes of evaporation, a constant contact radius (pinned) and a constant contact angle (receding) mode, allow for pinning-depinning transitions and model the evaporation dynamics accordingly using an evaporation description analogous to Eq.~\eqref{eq:evap}. Their results can be reproduced quantitatively by our model. However, our model also includes the relaxation of the droplet shape towards its equilibrium angle, after it is deposited on the surface with an angle different from the equilibrium value. We discuss in more detail the similarities and differences between their work and the results from our model in the next subsection.

\begin{figure}[htp!]
\centering
\includegraphics[width=\figwidth]{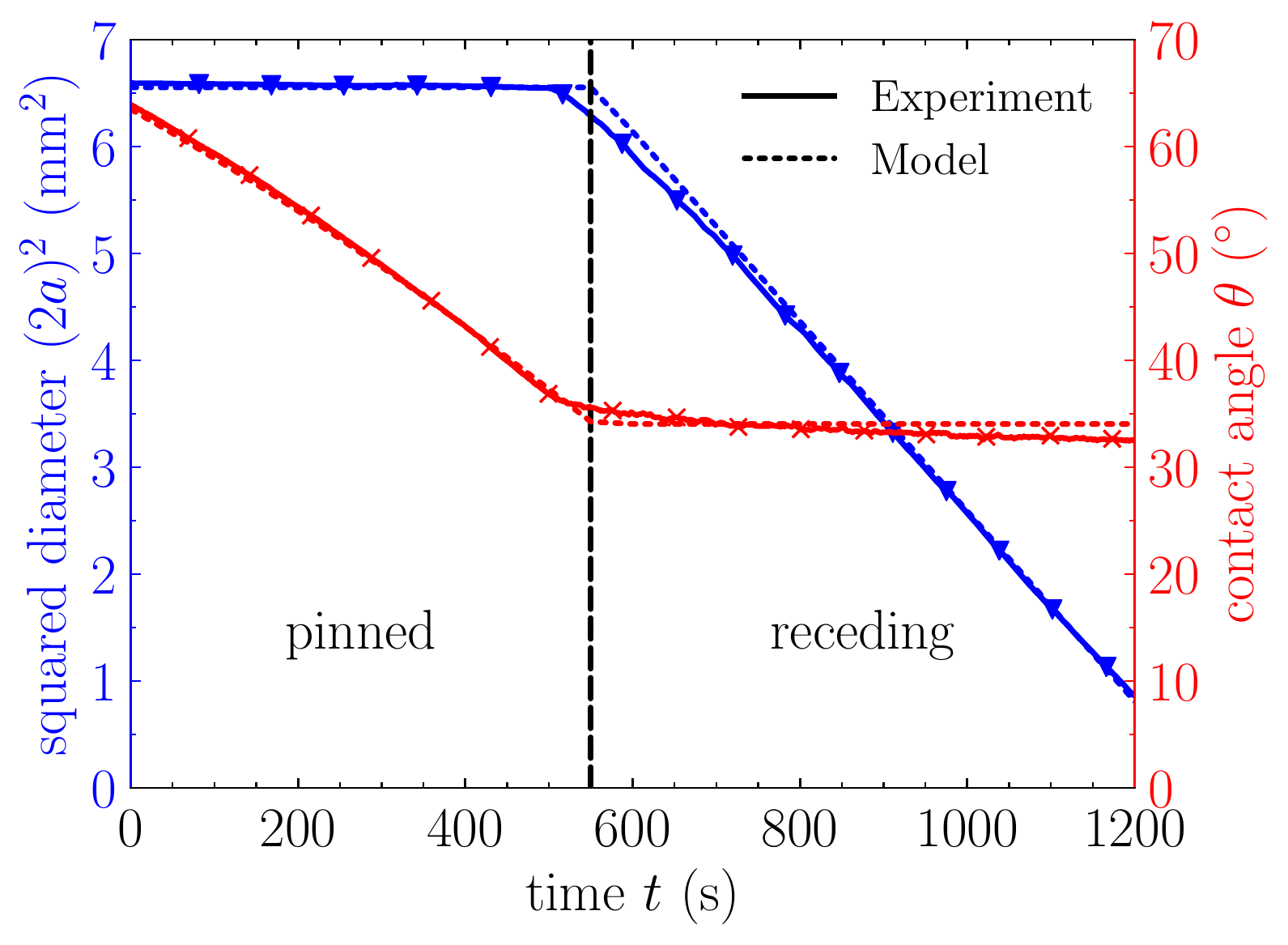}
\caption{Comparison between the results of experiments on the evaporation of sessile water droplets on a silica wafer~\cite{Belmiloud2012} (solid lines) and the numerical evaluation of the droplet model (dashed lines). The squared contact diameter ${(2a)^2}$ (left vertical axis, blue triangles) and contact angle ${\theta}$ (right vertical axis, red crosses) are shown as a function of time $t$. The two modes of evaporation, pinned and receding, are indicated. The model results are for a pinning force ${\fp\simeq \SI{0.034}{\newton\per\metre}}$, a vapour concentration difference ${\Delta c=\SI{11.6e-3}{\kilo\gram\per\meter\cubed}}$ and a vapour diffusion coefficient ${D=\SI{3.15e-5}{\metre\squared\per\second}}$.}
\label{fig:Belmiloud_compare}
\end{figure}

We now relate results from our model to the experimental data of \citet{Belmiloud2012} on the evaporation of a water droplet on a flat silicon surface, see Fig.~\ref{fig:Belmiloud_compare}. Figure~\ref{fig:Belmiloud_compare} shows the squared contact diameter $(2a)^2$ (blue triangles) and contact angle $\theta$ (red crosses) as a function of time ${t}$. The results of \citeauthor{Belmiloud2012}, represented by the solid lines, can be readily reproduced by our phenomenological model (dashed lines). Initially, the contact line of the droplet is pinned, as is seen from the squared contact diameter remaining constant, while the contact angle decreases. As the receding contact angle ${\thetar}$ is reached, a pinning-depinning transition occurs, after which the angle remains constant at the receding value and the diameter squared decreases linearly, as discussed in the \nameref{sec:theory} section.

In order to model the evaporation process of the initially pinned droplet, we choose our model parameters to correspond to the experimental values. The pinning force per unit length ${\fp}$ was set to a value of ${\fp\simeq \SI{0.034}{\newton\per\metre}}$ to obtain a receding contact angle of ${\thetar=\SI{34}{\degree}}$. The value for the pinning force per unit length is of the same order of magnitude as the liquid-air interfacial tension (${\gammaLG=\SI{0.07}{\newton\per\metre}}$). The values reported by~\citet{Belmiloud2012} for the surface vapour concentration ${c_s}$ and the relative humidity were used to determine ${\Delta c=\SI{11.6e-3}{\kilo\gram\per\meter\cubed}}$. The best correspondence between the measurement and our model is obtained for a vapour diffusion coefficient ${D=\SI{3.15e-5}{\metre\squared\per\second}}$, as opposed to the reported ${D=\SI{2.60e-5}{\metre\squared\per\second}}$. However, \citeauthor{Belmiloud2012} also report on an underestimation of the evaporation rate: the droplet evaporates faster than predicted by Eq.~\eqref{eq:evap}~\cite{Belmiloud2012}. This is arguably due to inaccuracies in measuring the properties of the ambient vapour.

\subsection{Predictions by full model}
\label{subsec:results_full_model}
We now consider the effect of the interplay between the three components of our model to predict the evaporation dynamics of a droplet. To that end, we first discuss two limiting cases. We report our findings on (1) the effect of contact line pinning on a non-evaporating, relaxing droplet, and (2) the effect of shape relaxation on the lifetimes of droplets with an unpinned contact line. Subsequently, we present our results on simultaneous shape relaxation and evaporation of a droplet subject to contact line pinning.

\begin{figure}[htp!]
\centering
\includegraphics[width=\figwidth]{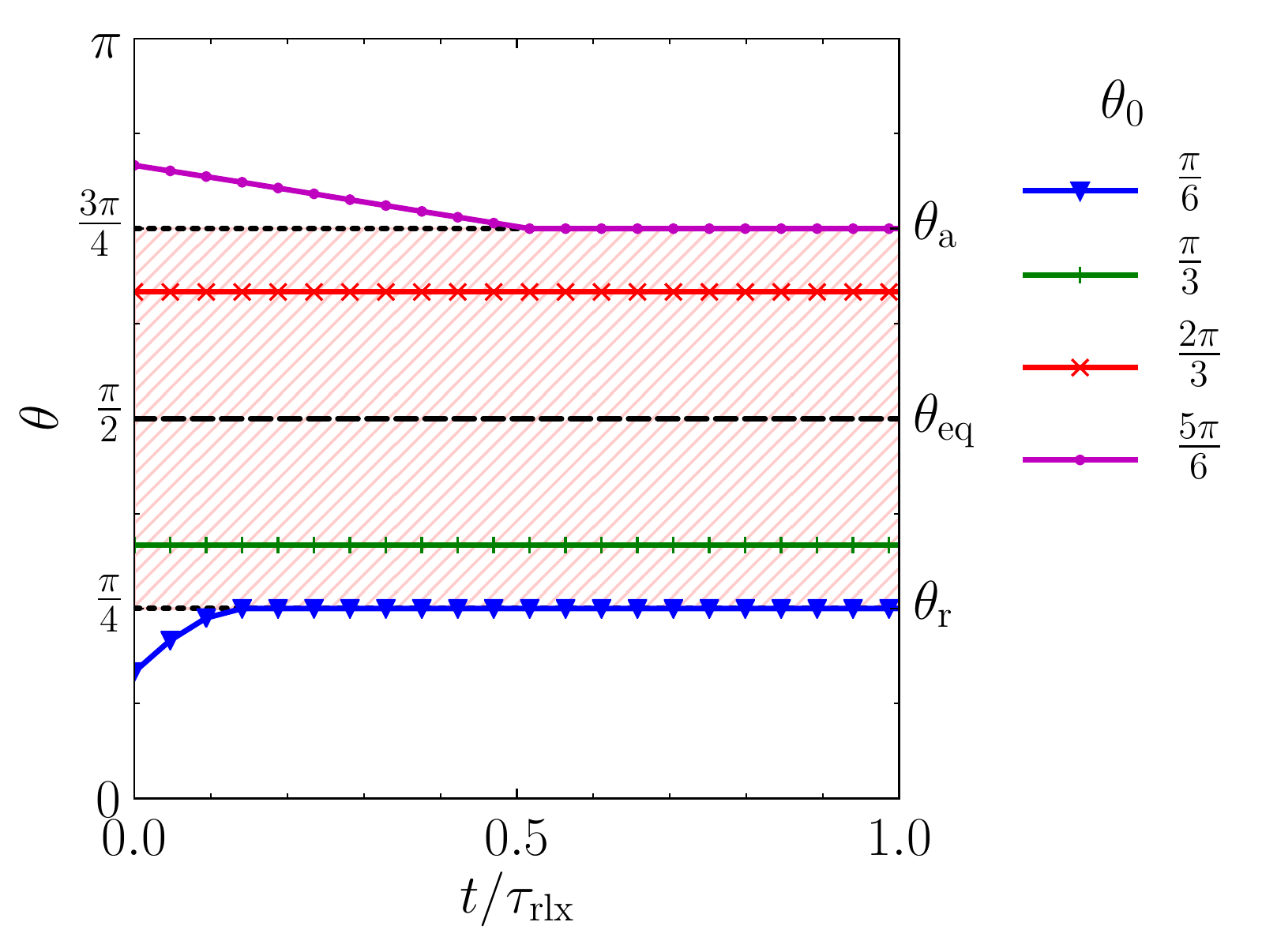}
\caption{Relaxation of the contact angle ${\theta}$ of a deposited drop on a surface towards the equilibrium value ${\thetaeq=\pi/2}$. There is no evaporation and the pinning force ${\fp}$ is set such that the receding and advancing contact angles are ${\thetar=\pi/4}$ and ${\thetaa=3\pi/4}$. This results in the fixed area range between ${\thetar}$ and ${\thetaa}$, where the capillary force ${\fc}$ cannot overcome ${\fp}$ and the contact line becomes or remains pinned.}
\label{fig:relaxation_pinning_thetaeq=90deg}
\end{figure}

If the shape relaxation of a droplet is affected by contact line pinning, the contact angle relaxation in the absence of evaporation studied in the previous section (Fig.~\ref{fig:relax_MKT_exp}) changes drastically, as is illustrated in Fig.~\ref{fig:relaxation_pinning_thetaeq=90deg}.
If droplets start out within the fixed area region, \latin{i.e.}, have an initial angle \mbox{${\thetar< \theta_0< \thetaa}$}, indicated by the shaded region in Fig.~\ref{fig:relaxation_pinning_thetaeq=90deg}, then the contact line is not able to move. In other words, the droplets are not able to relax their shape to accommodate the equilibrium contact angle ${\thetaeq}$. For initial angles outside of this regime, shape relaxation does occur, albeit only until the fixed area region is reached, after which the motion of the contact line is halted. This phenomenon has strong implications for the lifetime of an evaporating droplet. The asymmetry in the time it takes for the droplet to become pinned for ${\theta_0=\pi/6}$ and for ${\theta_0=5\pi/6}$ has its origin again, as is the case for the shape relaxation shown in Fig.~\ref{fig:relax_MKT_exp}, in the non-linearity of Eqs.~(\ref{eq:dyn_theta},\ref{eq:Gamma}).

If we allow for evaporation, the shape relaxation of a droplet from an initial contact angle ${\theta_0}$ towards its equilibrium angle ${\thetaeq}$ may have a strong impact on the evaporation dynamics of a droplet, also without any contact line pinning occurring. As discussed in the \nameref{sec:theory} section, the evaporation rate depends on the contact angle ${\theta}$, and is at its minimum for ${\theta={\pi}/{2}}$. If a droplet with a certain ${\thetaeq}$ is deposited onto a surface with an initial angle ${\theta_0\neq\thetaeq}$, the relative speeds at which the droplet relaxes to its equilibrium angle and at which it evaporates, characterised by the ratio ${\ratio}$, will influence the lifetime of such a droplet. In the remainder of our manuscript, we adopt the representation style of \citet{Stauber2014} when discussing lifetimes of droplets, where we depict the scaled evaporation time ${\tevap/\tauevap}$ as a function of the initial contact angle ${\theta_0}$. 

\begin{figure}[htp!]
\centering
\includegraphics[width=.9\figwidth]{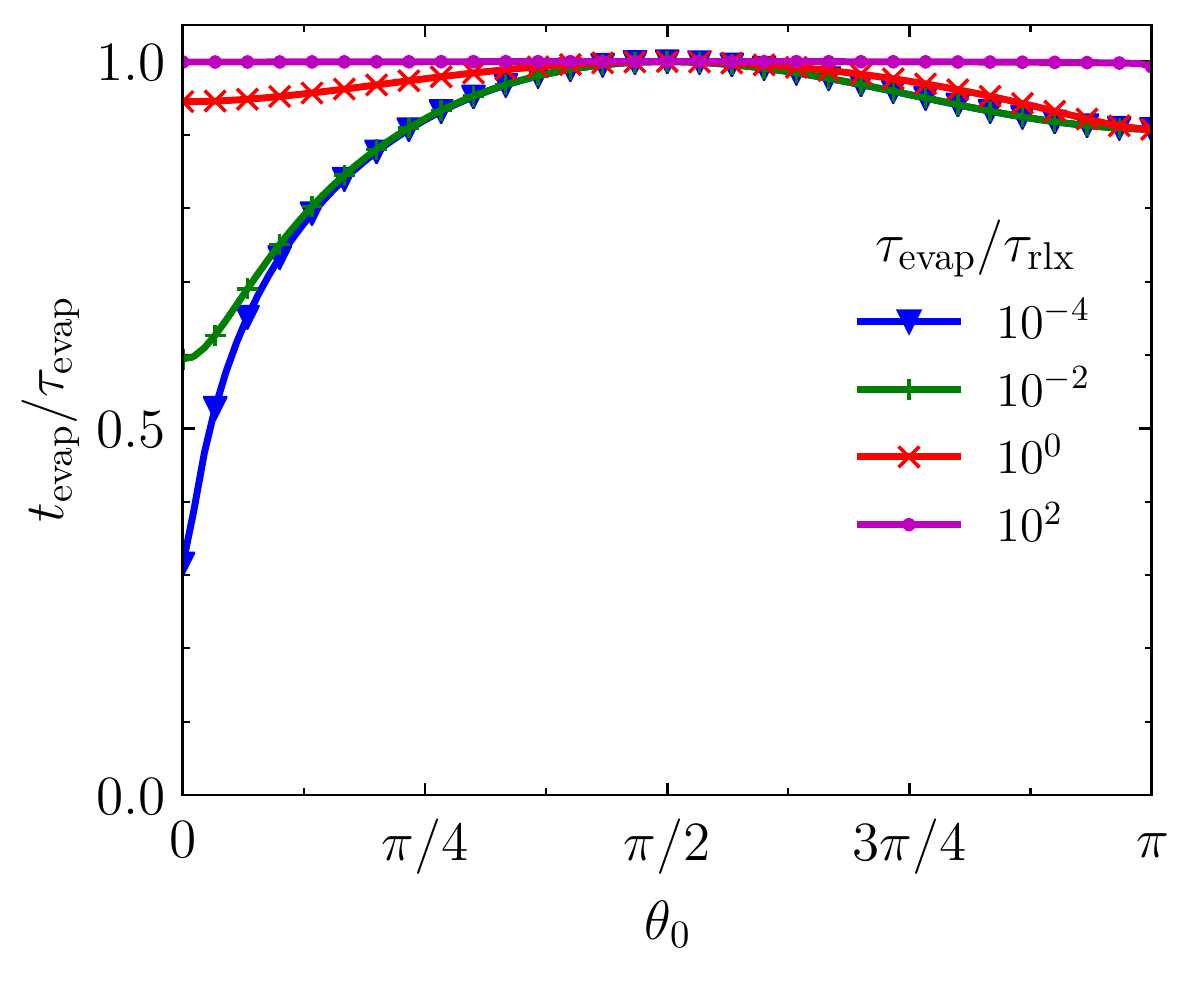}
\caption{Scaled evaporation times ${\tevap/\tauevap}$ for sessile droplets as a function of the initial contact angle ${\theta_0}$, for ratios ${\ratio=10^{-4},\ 10^{-2},\ 10^{0},\ 10^{2}}$. The equilibrium contact angle is ${\thetaeq=\pi/2}$.}
\label{fig:no_pinning_thetaeq=90deg}
\end{figure}

Firstly, we consider the evaporation of droplets in the absence of contact line pinning. In Fig.~\ref{fig:no_pinning_thetaeq=90deg} we present the droplet lifetimes ${\tevap/\tauevap}$ as a function of the initial contact angle ${\theta_0}$, for ${\ratio=10^{-4},\ 10^{-2},\ 10^{0},\ 10^{2}}$. In the limit of slow shape relaxation (${\ratio\ll 1}$), we exactly recover the result of \citet{Stauber2014} for evaporation with a constant contact angle, see the blue triangles in Fig.~\ref{fig:no_pinning_thetaeq=90deg}. The droplet lifetime decreases rapidly for $\theta_0\to 0$, as the area-to-volume ratio increases. As already discussed in the \nameref{sec:theory} section, the lifetime is longest for ${\theta=\pi/2}$, resulting in a maximum in the graph. For ${\theta>\pi/2}$ the lifetimes slightly decrease again due to the increase in area-to-volume ratio.

For increasing ${\ratio}$, we see a decreasing effect of the initial angle on the lifetimes, as the contact angles relax more quickly to the equilibrium value ${\thetaeq=\pi/2}$. The increase in droplet lifetime due to faster shape relaxation is most notable for small contact angles, as the relaxation is the fastest in that regime and small changes to the contact angle induce large changes in the evaporation rate, according to Eq.~\eqref{eq:evap}. Close to ${\theta_0=\pi/2}$, however, both shape relaxation and evaporation are slow and hence the changes in droplet lifetime upon changing ${\ratio}$ are small. For ${\tauevap/\taurlx\gg 1}$, the droplet lifetime is effectively independent of $\theta_0$. We conclude that the evaporation dynamics of a droplet on a flat surface is strongly affected by the ratio between the rates of evaporation and shape relaxation, in the absence of contact line pinning.

If the contact line can become stuck on the surface due to pinning, however, the response of a drop to deposition on a surface becomes more complex. We can identify three regimes in the droplet dynamics, distinguished by the value of the initial contact angle ${\theta_0}$ relative to the receding and advancing contact angles ${\thetar}$ and ${\thetaa}$: 
\begin{enumerate}[a.]
\item $\theta_0 < \thetar$, where $\theta_0$ lies below the fixed area range;
\item $\thetar \leq \theta_0 \leq \thetaa$, where $\theta_0$ lies within the fixed area range;
\item $\theta_0 > \thetaa$, where $\theta_0$ lies above the fixed area range.
\end{enumerate}

\begin{figure}[htp!]
\centering
\includegraphics[width=\figwidth]{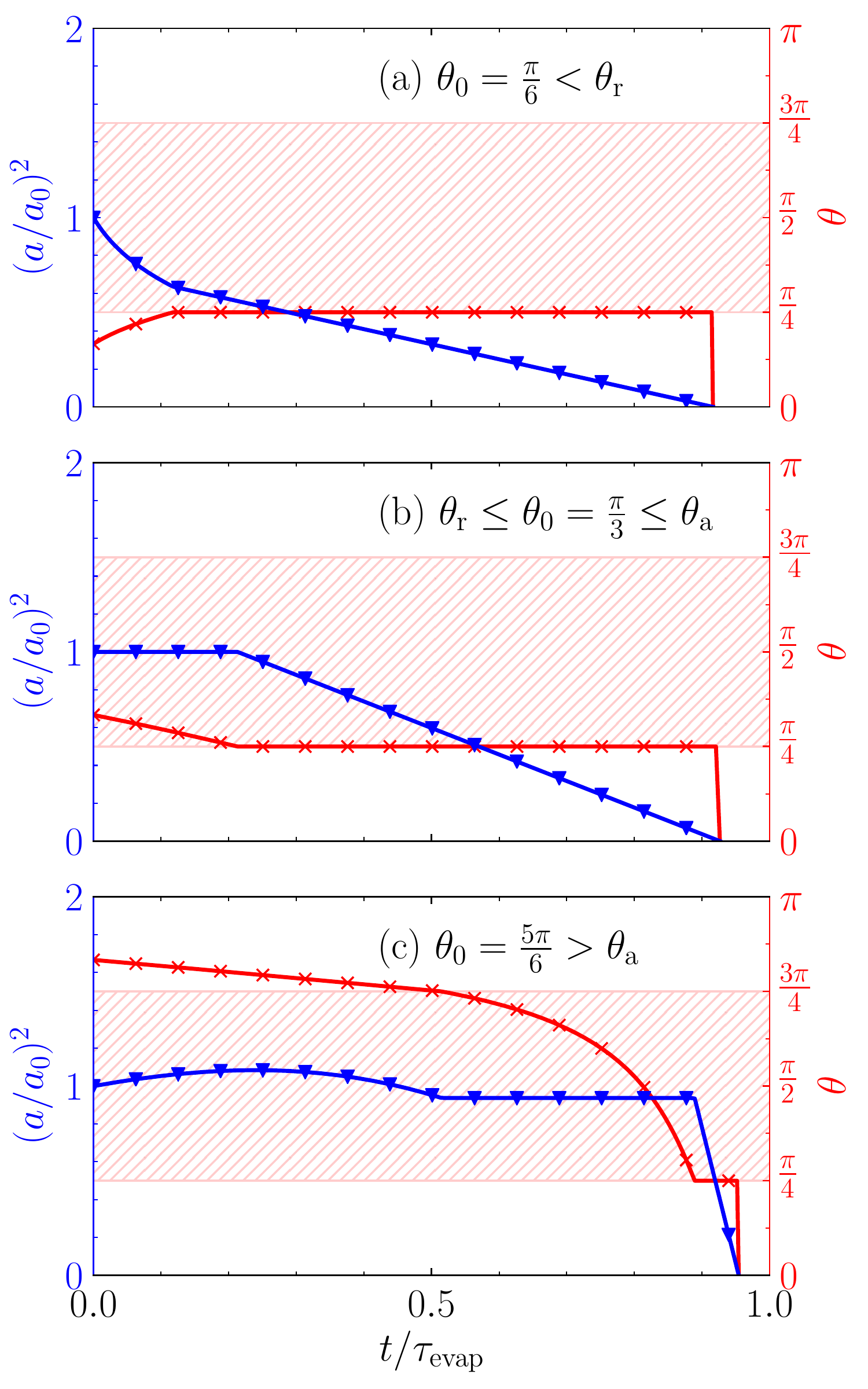}
\caption{Scaled contact area ${(a/a_0)^2}$ and contact angle ${\theta}$ as a function of non-dimensional time ${t/\tauevap}$, for the evaporation of a droplet with equilibrium angle ${\thetaeq=\pi/2}$. The shaded areas indicate the fixed area domains between the receding and advancing contact angles, ${\thetar=\pi/4}$ and ${\thetaa=3\pi/4}$. The ratio between evaporation and relaxation timescales is set to ${\ratio=1}$, and we show the time evolution of the droplet shape for three values of the initial contact angle ${\theta_0}$, (a)~${\theta_0=\pi/6<\thetar}$, (b)~${\thetar<\theta_0=\pi/3<\thetaa}$, (c)~${\theta_0=5\pi/6>\thetaa}$.}
\label{fig:area_theta_relax_pinning}
\end{figure}

Fig.~\ref{fig:area_theta_relax_pinning} shows the droplet shape in terms of the scaled squared radius ${(a/a_0)^2}$ (blue triangles) and ${\theta}$ (red crosses) as a function of non-dimensional time ${t/\tauevap}$. The pinning force ${\fp}$ was set to ${\fp\approx 0.924\,\gammaLG}$ leading to ${\thetar=\pi/4}$ and ${\thetaa=3\pi/4}$. The contact angle range between the two (the fixed area domain) is indicated by the shaded area. The ratio between evaporation and relaxation timescales is ${\ratio=1}$.
Three different graphs are shown to illustrate the three regimes. If ${\theta_0<\thetar}$, the droplet starts out in the depinned state, and thus the contact line moves to accommodate the contact angle relaxation towards the equilibrium value ${\thetaeq}$, see Fig.~\ref{fig:area_theta_relax_pinning}(a). After some time, the receding contact angle ${\thetar}$ is reached and the contact angle relaxation halts, resulting in the remainder of the evaporation process occurring with a constant contact angle $\thetar$. In the second case, shown in Fig.~\ref{fig:area_theta_relax_pinning}(b), the droplet starts out with a pinned contact line, since ${\thetar<\theta_0<\thetaa}$. Due to evaporation, the contact angle decreases until it reaches ${\thetar}$. At that point, a depinning transition occurs, the contact line is allowed to move again and evaporation continues with a constant contact angle $\thetar$. Parenthetically, in Fig.~\ref{fig:area_theta_relax_pinning}(b), we present the results for an initial contact angle $\theta_0<\pi/2$, but the general behaviour of the contact angle as a function of time is the same for ${\pi/2\leq\theta_0<\thetaa}$. Lastly, for ${\theta_0>\thetaa}$, the contact line can initially move freely, causing the droplet to spread and increase its contact area, see Fig.~\ref{fig:area_theta_relax_pinning}(c). After the contact angle reaches the advancing value ${\thetaa}$, however, the contact line becomes pinned. Once more, from this point on, the contact angle decreases due to evaporation. And again, as the contact angle reaches the receding value $\thetar$, a depinning transition occurs and evaporation continues with a constant contact angle.

\begin{figure}[htp!]
\centering
\includegraphics[width=0.82\figwidth]{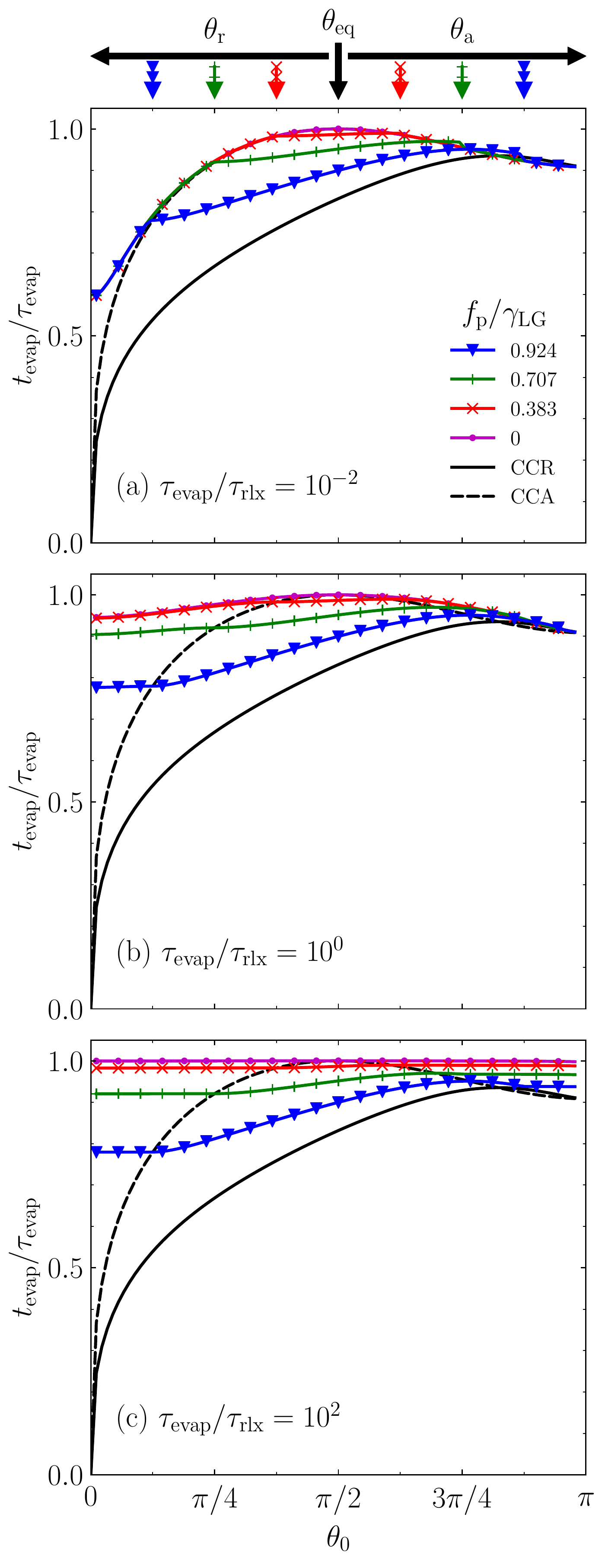}
\caption{Scaled evaporation times ${\tevap/\tauevap}$ of sessile droplets as a function of the initial contact angle ${\theta_0}$, for various values for ${\fp/\gammaLG\approx 0.924,\ 0.707,\ 0.383,\ 0}$, and three ratios ${\ratio}$ of (a)~${10^{-2}}$, (b)~${10^0}$, (c)~${10^2}$, covering the range from fast to slow evaporation. The values of the respective receding and advancing contact angles are indicated by arrows in the same colour as the corresponding lines in the graph. The black lines represent the two limiting cases of evaporation at constant radius of the contact area (CCR, solid) and evaporation at a constant contact angle (CCA, dashed). The equilibrium contact angle is ${\thetaeq=\pi/2}$.}
\label{fig:relax_pinning}
\end{figure}

The emergence of the three regimes due to the presence of contact line pinning has a significant impact on the droplet lifetimes. The extent of the effect depends on the ratio of evaporation and relaxation timescales ${\ratio}$, as is illustrated in Fig.~\ref{fig:relax_pinning}.
Figure~\ref{fig:relax_pinning} shows the scaled lifetime of an evaporating droplet ${\tevap/\tauevap}$ as a function of the initial contact angle ${\theta_0}$, for ${\ratio=10^{-2},\ 10^0,\ 10^2}$, covering the entire range from fast to slow evaporation. We can identify two limiting cases for the evaporation in all three graphs, shown in black, being evaporation with a constant contact radius (CCR, solid lines) and evaporation with a constant contact angle (CCA, dashed lines). These two limits are not dependent on the ratio $\ratio$, as we impose that either the contact area or the constant angle remains fixed.

The time it takes for a droplet to evaporate is shorter at a constant radius compared to a constant angle for the majority of the initial angle range ${0\leq \theta_0 \leq \pi}$. This is because the constant contact radius mode causes the contact angle to decrease during evaporation. Decreasing $\theta$ generally speeds up the evaporation process due to an increase of the surface-to-volume ratio, especially at late times. For large initial angles (${\theta_0\to\pi}$), however,  evaporation in the constant angle mode becomes faster than evaporation in the constant radius mode. As the latter causes a continuous decrease in the contact angle, it initially slows down the evaporation rate before speeding it up again. As discussed before, the maximum lifetime of a droplet evaporating in the constant angle mode is ${\tevap=\tauevap}$ for ${\theta_0=\pi/2}$, resulting in a maximum in the graph. For the constant radius mode, the maximum lifetime is shorter than the maximum in the constant angle mode, being ${\tevap\approx 0.9354\,\tauevap}$ for ${\theta_0\approx 0.822\pi}$~\cite{Stauber2014}.

The differently coloured arrows at the top of Fig.~\ref{fig:relax_pinning} represent the receding contact angle $\thetar$ and the advancing contact angle $\thetaa$. These define the domain in which the contact area remains constant. The arrow colours correspond to the colours of the curves shown in the figure, which depict the droplet lifetimes $\tevap/\tauevap$ as a function of the initial contact angle $\theta_0$, for  values of the pinning force $\fp/\gammaLG$ of approximately 0.924 (blue triangles), 0.707 (green pluses), 0.383 (red crosses) and 0 (purple dots). The values for $\thetar$ and $\thetaa$ remain constant in all three graphs, and as $\thetaeq=\pi/2$, they take symmetric values around the equilibrium.

For all three graphs in Figs.~\ref{fig:relax_pinning}(a) to (c), the curve segments between the bounding receding and advancing contact angles are identical. The reason is that if the contact line of a drop is initially pinned due to the choice of initial angle, this angle cannot relax towards its equilibrium value as is also shown in Fig.~\ref{fig:area_theta_relax_pinning}(b). Therefore, the magnitude of the shape relaxation rate does not affect the evaporation process. During evaporation, the contact angle only decreases from the initial to the receding angle and remains at that value until the drop has fully evaporated. 

The relaxation of the contact angle towards its equilibrium value is only possible for initial angles outside of the fixed area domain, as shown in Fig.~\ref{fig:area_theta_relax_pinning}(a) and (c), where shape relaxation occurs until the contact angle reaches either boundary. In other words, for values of $\theta_0$ outside of the fixed area region, the relative shape relaxation rate does have an impact on the droplet lifetime. For values of $\ratio\gtrsim 1$, as shown in Fig.~\ref{fig:relax_pinning}(b) and (c), evaporation is relatively slow and relaxation in essence instantaneous. This leads to an evaporation time that is essentially an invariant of the initial angle, outside of the fixed area domain, where the lifetime takes on the value at the nearest boundary (at $\thetar$ or $\thetaa$). If evaporation is very fast, \latin{i.e.}, $\ratio\ll 1$, shown in Fig.~\ref{fig:relax_pinning}(a), relaxation cannot keep up and the evaporation time is dictated by a virtually constant contact angle. For sufficiently small $\theta_0$, however, relaxation can keep up with evaporation and the evaporation time deviates from the lifetimes for the constant contact angle mode. This deviation vanishes for $\ratio\to 0$. 

For initial angles above the advancing angle, the lifetime curves start to deviate from both limiting cases and from the curves reported by \citet{Stauber2014}, when ${\ratio}$ increases. This is caused by the evaporation dynamics predicted by our model being more complicated than a simple imposed transition from a pinned into a depinned state. As shown above in Fig.~\ref{fig:area_theta_relax_pinning}(c), the contact line can initially move freely, implying that the contact angle starts to move towards its equilibrium value $\thetaeq$. Upon reaching $\thetaa$, the contact line becomes pinned and the contact angle decreases until it reaches $\thetar$. Subsequently, a pinning-depinning transition occurs and the droplet evaporates with a fixed contact angle. In other words, the droplet experiences two transitions, rather than one, by subsequently going through depinned, pinned and depinned modes.

\section{Discussion}
\label{sec:discussion}
We developed a phenomenological model for the shape relaxation of an evaporating droplet. The characteristic time scale associated with this relaxation is found to be proportional to a length scale $L$. This length scale $L$ has been connected to (1) a slip or friction length, or (2) the size of the droplet. In order to incorporate the effects of either length scale on the evaporation dynamics, we have equipped Eq.~\eqref{eq:Gamma}, which describes the droplet relaxation, with a scale factor $\alpha$. In the discussion of the results in the previous section we chose $\alpha=1$ for simplicity. 

Now we discuss in more detail the implications of considering an alternative $\alpha=\alpha(t)$, which is proportional to the droplet size:
\begin{equation}
\alpha(t) = \left[\frac{V(t)}{V_0}\right]^{1/3},
\end{equation}
where $V(t)$ and $V_0$ denote the instantaneous and initial droplet volumes, respectively. As a consequence, the relaxation process speeds up as the droplet size decreases. We find, however, that explicitly taking this effect into account hardly affects the droplet lifetime. This is caused by the circumstance that the capillary driving force is the strongest at short times, as the difference $\cos\theta-\cos\thetaeq$ is then the greatest. In other words, the majority of the relaxation process occurs at short times. However, at short times the droplet has hardly lost any volume by evaporation, which means that the scale factor $\alpha(t)\approx 1$, causing the relaxation processes for both expressions for $\alpha$ to occur in virtually the same manner.
\begin{figure}[htp!]
\centering
\includegraphics[width=.95\figwidth]{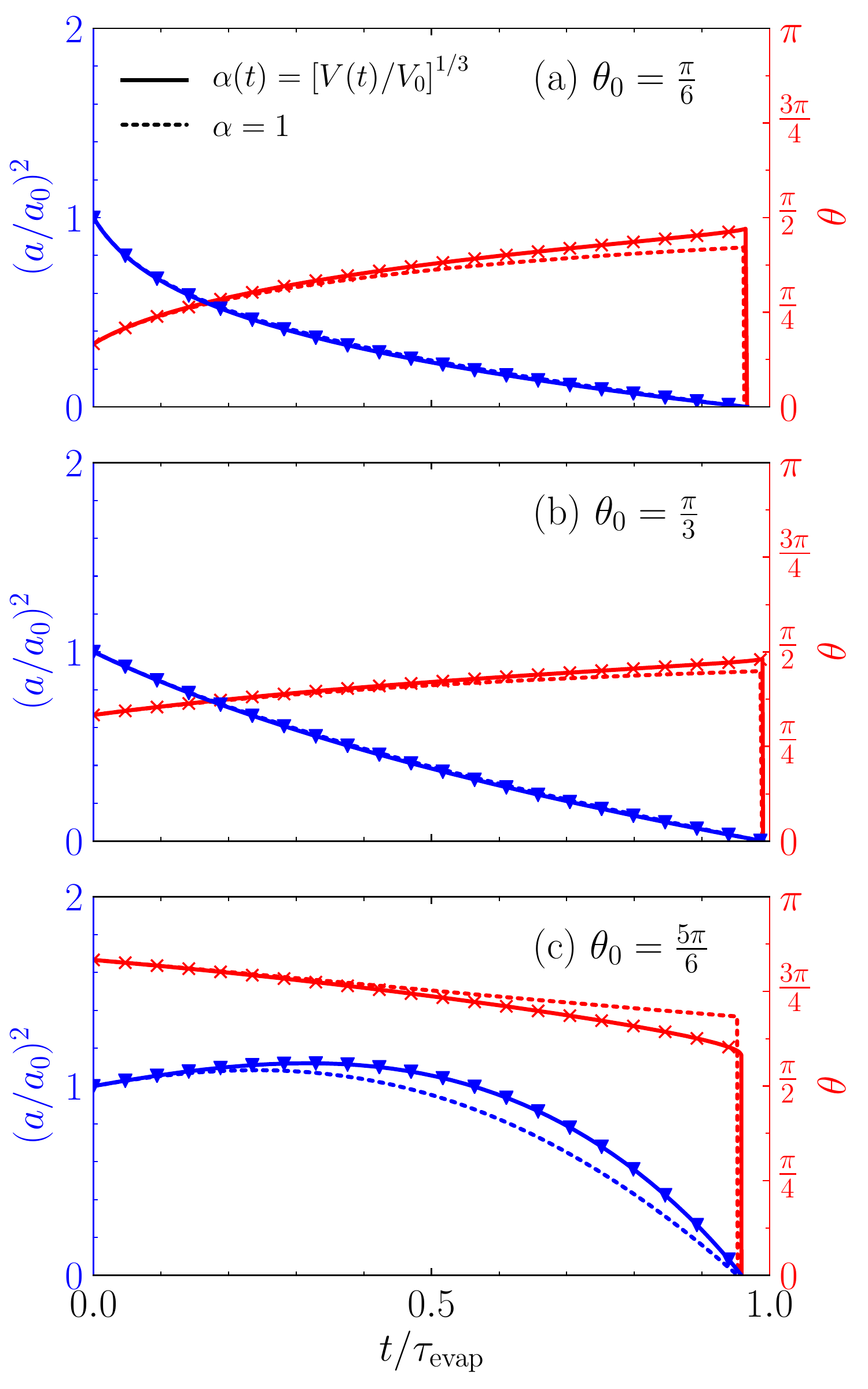}
\caption{Comparison of the squared scaled radius ${(a/a_0)^2}$ and the contact angle ${\theta}$ between the alternative scale factor ${\alpha(t)=\left[V(t)/V_0\right]^{1/3}}$ (solid lines) and the original ${\alpha=1}$ (dashed lines), as a function of scaled time ${t/\tauevap}$. ${\ratio=1}$ and figures are shown for initial angles (a)~${\theta_0=\pi/6}$, (b)~${\theta_0=\pi/3}$, (c)~${\theta_0=5\pi/6}$.}
\label{fig:area_theta_stack_alttau}
\end{figure}
In Fig.~\ref{fig:area_theta_stack_alttau} we depict the squared scaled radius $(a/a_0)^2$ (blue triangles) and the contact angle $\theta$ (red crosses) of evaporating droplets as a function of scaled time $t/\tauevap$. We present the results for the scale factor ${\alpha(t)=[V(t)/V_0]^{1/3}}$ (solid lines) compared to ${\alpha=1}$ (dashed lines), for three values of $\theta_0$. For these calculations, we do not incorporate a contact line pinning force, which means the droplet is allowed to relax its shape towards $\thetaeq=\pi/2$, and ${\ratio=1}$.

In all three graphs, we can clearly see that the dynamics described by the two expressions for $\alpha$ are identical at early times. Only after the droplet has partly evaporated, we see a slight deviation in the dynamics, due to the decrease of $\alpha(t)$. This effect only arises after approximately 30\% of the evaporation time has passed. For large initial contact angles $\theta_0$, which we show in Fig.~\ref{fig:area_theta_stack_alttau}(c), the deviations between the graphs for the two expressions for $\alpha$ are slightly larger than for smaller $\theta_0$ (Fig.~\ref{fig:area_theta_stack_alttau}(a) and (b)). However, the time at which the droplet is fully evaporated is hardly affected. Note that the presence of contact line pinning would only decrease the effect shape relaxation has on the evaporation time, as it inhibits contact line motion for a certain range of the contact angle $\theta$. We conclude from this that the lifetime of an evaporating droplet is not sensitive to our choice for $\alpha$ and hence explicitly taking into account the size-dependence of the relaxation process has a negligible effect on the total evaporation time.

A second point left for discussion is the influence of the value of the equilibrium angle $\thetaeq$ on the evaporation dynamics of a droplet. We recall that an equilibrium contact angle of $\pi/2$ has two implications for the evaporation dynamics of a droplet. Firstly, evaporation is slowest at $\theta=\pi/2$, as described by Eq.~\eqref{eq:evap}. Secondly, in the presence of contact line pinning, the fixed area domain is located symmetrically around the equilibrium angle: as ${\cos\thetaeq=0}$, the advancing contact angle is given by ${\thetaa=\pi-\thetar}$, as described by Eq.~\eqref{eq:theta_r} and \eqref{eq:theta_a}. Both properties are not valid for droplets with an equilibrium contact angle ${\thetaeq\neq \pi/2}$. A droplet taking on a non-hemispherical equilibrium results on the one hand in the fact that faster shape relaxation, or increasing $\ratio$, not necessarily implies slower evaporation, as we have shown in Fig.~\ref{fig:relax_pinning}, but that this depends on the initial and equilibrium contact angles. On the other hand, the receding and advancing contact angles are now located asymmetrically around the equilibrium angle.

As we have seen in the \nameref{sec:results} section, if contact line pinning occurs, the time it takes for a droplet to evaporate depends strongly on the values of the receding and advancing contact angles (see Fig.~\ref{fig:relax_pinning}). For contact angles in between the two, contact line motion is inhibited. For ${\thetaeq\neq \pi/2}$, this principle is still valid, only $\thetar$ and $\thetaa$ are located asymmetrically around $\thetaeq$. For initial angles within the fixed area domain, the lifetimes remain unchanged with respect to $\thetaeq=\pi/2$. As contact line motion is inhibited there, shape relaxation is blocked, so the value of $\thetaeq$ is irrelevant. For initial angles outside of the domain the lifetime as a function of initial angle behaves similar to what we show in Fig.~\ref{fig:relax_pinning}: for increasing shape relaxation rates, so increasing $\ratio$, the evaporation time of a droplet converges to the values at the boundaries and it becomes increasingly less dependent on the initial contact angle $\theta_0$.

We conclude from the above that the specific value of the equilibrium contact angle $\thetaeq$ has little effect on the general behaviour of the droplet lifetime as a function of the initial contact angle. It does affect the evaporation dynamics, but to an extent that is limited to two factors. On the one hand, it determines the evaporation time at the equilibrium angle, so it affects the droplet lifetime most in the absence of contact line pinning and in the limit of fast shape relaxation. On the other hand, it determines, together with the magnitude of the pinning force $\fp$, the locations of the receding and advancing contact angles, which in turn define the region in which contact line motion is inhibited. 

\section{Summary and Conclusions}
\label{sec:conclusion}
In conclusion, we propose a model for diffusive evaporation of a droplet on a flat surface, which accounts for the relaxation of the contact angle towards its equilibrium value. This shape relaxation is driven by the tendency of the droplet to reach its minimum free energy state. We also model pinning of the contact line onto the surface by introducing a pinning force, insisting that the contact line remains pinned as long as the capillary forces are not able to overcome this threshold force.

Within our model description, the time it takes for such a droplet to evaporate turns out to depend on five parameters, being the initial and equilibrium contact angles, the characteristic timescales associated with shape relaxation and evaporation, and the magnitude of the contact line pinning force. The ratio between the two characteristic timescales describes the competition between shape relaxation and evaporation, which has a significant effect on the droplets' lifetime. In the limit of slow relaxation (or fast evaporation), the total evaporation time of a droplet strongly depends on the initial contact angle, whereas for fast relaxation the lifetime is virtually  unaffected by the value of the initial contact angle.

The presence of a pinning force results in a contact angle range for which the contact line is fixed, as the capillary forces are not capable of overcoming the pinning force. This regime is bounded by the receding and advancing contact angles and as long as the contact angle resides within this range, the contact area remains constant. The magnitude of the pinning force determines the values of the receding and advancing contact angles and therefore has an impact on the lifetime of an evaporating droplet: the shape relaxation of a droplet becomes partly suppressed, because the droplet cannot relax its shape for contact angles within this fixed area regime.

We show that shape relaxation has a significant impact on the evaporation time of a droplet, both in the absence and presence of contact line pinning. Explicitly taking into account the size-dependence of the relaxation process turns out to have virtually no effect on the droplet's lifetime, since the majority of the relaxation occurs at short times for which the droplet size has hardly decreased.  The value of $\thetaeq$ does also not affect the general dynamical behaviour, however, it does define the lifetime for a droplet at its equilibrium angle and the location of the receding and advancing contact angles.

Finally, the simplicity of our model allows for relatively straightforward evaluation of the dynamics of an evaporating droplet. This means that it can also be readily extended to, \latin{e.g.}, take into account compound exchange between the solid and the liquid phase, or investigate an evaporation process wherein the droplet properties do not remain constant in time.

\begin{acknowledgement}
This work is part of the research programme with project number 13919, which is partly financed by the Netherlands Organisation for Scientific Research (NWO). We thank Theo Polet, Cor Rops, Gijsbert Rispens and Daan van Sommeren for valuable discussions.
\end{acknowledgement}

\bibliography{literature}

\providecommand{\latin}[1]{#1}
\makeatletter
\providecommand{\doi}
  {\begingroup\let\do\@makeother\dospecials
  \catcode`\{=1 \catcode`\}=2 \doi@aux}
\providecommand{\doi@aux}[1]{\endgroup\texttt{#1}}
\makeatother
\providecommand*\mcitethebibliography{\thebibliography}
\csname @ifundefined\endcsname{endmcitethebibliography}
  {\let\endmcitethebibliography\endthebibliography}{}
\begin{mcitethebibliography}{58}
\providecommand*\natexlab[1]{#1}
\providecommand*\mciteSetBstSublistMode[1]{}
\providecommand*\mciteSetBstMaxWidthForm[2]{}
\providecommand*\mciteBstWouldAddEndPuncttrue
  {\def\EndOfBibitem{\unskip.}}
\providecommand*\mciteBstWouldAddEndPunctfalse
  {\let\EndOfBibitem\relax}
\providecommand*\mciteSetBstMidEndSepPunct[3]{}
\providecommand*\mciteSetBstSublistLabelBeginEnd[3]{}
\providecommand*\EndOfBibitem{}
\mciteSetBstSublistMode{f}
\mciteSetBstMaxWidthForm{subitem}{(\alph{mcitesubitemcount})}
\mciteSetBstSublistLabelBeginEnd
  {\mcitemaxwidthsubitemform\space}
  {\relax}
  {\relax}

\bibitem[Kuang \latin{et~al.}(2014)Kuang, Wang, and Song]{Kuang2014}
Kuang,~M.; Wang,~L.; Song,~Y. {Controllable Printing Droplets for
  High-Resolution Patterns}. \emph{Adv. Mater.} \textbf{2014}, \emph{26},
  6950--6958\relax
\mciteBstWouldAddEndPuncttrue
\mciteSetBstMidEndSepPunct{\mcitedefaultmidpunct}
{\mcitedefaultendpunct}{\mcitedefaultseppunct}\relax
\EndOfBibitem
\bibitem[Park and Moon(2006)Park, and Moon]{Park2006}
Park,~J.; Moon,~J. {Control of Colloidal Particle Deposit Patterns within
  Picoliter Droplets Ejected by Ink-Jet Printing}. \emph{Langmuir}
  \textbf{2006}, \emph{22}, 3506--3513\relax
\mciteBstWouldAddEndPuncttrue
\mciteSetBstMidEndSepPunct{\mcitedefaultmidpunct}
{\mcitedefaultendpunct}{\mcitedefaultseppunct}\relax
\EndOfBibitem
\bibitem[Calvert(2001)]{Calvert2001}
Calvert,~P. {Inkjet Printing for Materials and Devices}. \emph{Chem. Mater.}
  \textbf{2001}, \emph{13}, 3299--3305\relax
\mciteBstWouldAddEndPuncttrue
\mciteSetBstMidEndSepPunct{\mcitedefaultmidpunct}
{\mcitedefaultendpunct}{\mcitedefaultseppunct}\relax
\EndOfBibitem
\bibitem[Yu \latin{et~al.}(2009)Yu, Zhu, Frantz, Reding, Chan, and
  Ozkan]{Yu2009}
Yu,~Y.; Zhu,~H.; Frantz,~J.~M.; Reding,~M.~E.; Chan,~K.~C.; Ozkan,~H.~E.
  {Evaporation and coverage area of pesticide droplets on hairy and waxy
  leaves}. \emph{Biosyst. Eng.} \textbf{2009}, \emph{104}, 324--334\relax
\mciteBstWouldAddEndPuncttrue
\mciteSetBstMidEndSepPunct{\mcitedefaultmidpunct}
{\mcitedefaultendpunct}{\mcitedefaultseppunct}\relax
\EndOfBibitem
\bibitem[Wei and Brainard(2009)Wei, and Brainard]{Wei2009}
Wei,~Y.; Brainard,~R.~L. \emph{{Advanced Processes for 193-nm Immersion
  Lithography}}; SPIE: Bellingham, 2009\relax
\mciteBstWouldAddEndPuncttrue
\mciteSetBstMidEndSepPunct{\mcitedefaultmidpunct}
{\mcitedefaultendpunct}{\mcitedefaultseppunct}\relax
\EndOfBibitem
\bibitem[Belmiloud \latin{et~al.}(2012)Belmiloud, Tamaddon, Mertens, Struyf,
  and Xu]{Belmiloud2012}
Belmiloud,~N.; Tamaddon,~A.~H.; Mertens,~P.~W.; Struyf,~H.; Xu,~X. {Dynamics of
  the Drying Defects Left by Residual Ultra-Pure Water Droplets on Silicon
  Substrate}. \emph{ECS J. Solid State Sci. Technol.} \textbf{2012}, \emph{1},
  P34--P39\relax
\mciteBstWouldAddEndPuncttrue
\mciteSetBstMidEndSepPunct{\mcitedefaultmidpunct}
{\mcitedefaultendpunct}{\mcitedefaultseppunct}\relax
\EndOfBibitem
\bibitem[Mack(2007)]{Mack2007}
Mack,~C. \emph{{Fundamental Principles of Optical Lithography}}; John Wiley
  {\&} Sons, Ltd: Chichester, UK, 2007\relax
\mciteBstWouldAddEndPuncttrue
\mciteSetBstMidEndSepPunct{\mcitedefaultmidpunct}
{\mcitedefaultendpunct}{\mcitedefaultseppunct}\relax
\EndOfBibitem
\bibitem[Lin(1987)]{Lin1987}
Lin,~B.~J. {The future of subhalf-micrometer optical lithography}.
  \emph{Microelectron. Eng.} \textbf{1987}, \emph{6}, 31--51\relax
\mciteBstWouldAddEndPuncttrue
\mciteSetBstMidEndSepPunct{\mcitedefaultmidpunct}
{\mcitedefaultendpunct}{\mcitedefaultseppunct}\relax
\EndOfBibitem
\bibitem[Kocsis \latin{et~al.}(2006)Kocsis, {Van Den Heuvel}, Gronheid,
  Maenhoudt, Vangoidsenhoven, Wells, Stepanenko, Benndorf, Kim, Kishimura,
  Ercken, {Van Roey}, O'Brien, Fyen, Foubert, Moerman, and
  Streefkerk]{Kocsis2006}
Kocsis,~M. \latin{et~al.}  {Immersion specific defect mechanisms: findings and
  recommendations for their control}. Proc. SPIE 6154, Opt. Microlithogr. XIX.
  2006; p 615409\relax
\mciteBstWouldAddEndPuncttrue
\mciteSetBstMidEndSepPunct{\mcitedefaultmidpunct}
{\mcitedefaultendpunct}{\mcitedefaultseppunct}\relax
\EndOfBibitem
\bibitem[Bourg{\`{e}}s-Monnier and Shanahan(1995)Bourg{\`{e}}s-Monnier, and
  Shanahan]{Bourges-Monnier1995}
Bourg{\`{e}}s-Monnier,~C.; Shanahan,~M. E.~R. {Influence of Evaporation on
  Contact Angle}. \emph{Langmuir} \textbf{1995}, \emph{11}, 2820--2829\relax
\mciteBstWouldAddEndPuncttrue
\mciteSetBstMidEndSepPunct{\mcitedefaultmidpunct}
{\mcitedefaultendpunct}{\mcitedefaultseppunct}\relax
\EndOfBibitem
\bibitem[Uno \latin{et~al.}(1998)Uno, Hayashi, Hayashi, Ito, and
  Kitano]{Uno1998}
Uno,~K.; Hayashi,~K.; Hayashi,~T.; Ito,~K.; Kitano,~H. {Particle adsorption in
  evaporating droplets of polymer latex dispersions on hydrophilic and
  hydrophobic surfaces}. \emph{Colloid Polym. Sci.} \textbf{1998}, \emph{276},
  810--815\relax
\mciteBstWouldAddEndPuncttrue
\mciteSetBstMidEndSepPunct{\mcitedefaultmidpunct}
{\mcitedefaultendpunct}{\mcitedefaultseppunct}\relax
\EndOfBibitem
\bibitem[Cachile \latin{et~al.}(2002)Cachile, B{\'{e}}nichou, Poulard, and
  Cazabat]{Cachile2002}
Cachile,~M.; B{\'{e}}nichou,~O.; Poulard,~C.; Cazabat,~A.~M. {Evaporating
  Droplets}. \emph{Langmuir} \textbf{2002}, \emph{18}, 8070--8078\relax
\mciteBstWouldAddEndPuncttrue
\mciteSetBstMidEndSepPunct{\mcitedefaultmidpunct}
{\mcitedefaultendpunct}{\mcitedefaultseppunct}\relax
\EndOfBibitem
\bibitem[Fukai \latin{et~al.}(2006)Fukai, Ishizuka, Sakai, Kaneda, Morita, and
  Takahara]{Fukai2006}
Fukai,~J.; Ishizuka,~H.; Sakai,~Y.; Kaneda,~M.; Morita,~M.; Takahara,~A.
  {Effects of droplet size and solute concentration on drying process of
  polymer solution droplets deposited on homogeneous surfaces}. \emph{Int. J.
  Heat Mass Transf.} \textbf{2006}, \emph{49}, 3561--3567\relax
\mciteBstWouldAddEndPuncttrue
\mciteSetBstMidEndSepPunct{\mcitedefaultmidpunct}
{\mcitedefaultendpunct}{\mcitedefaultseppunct}\relax
\EndOfBibitem
\bibitem[Yu \latin{et~al.}(2012)Yu, Wang, and Zhao]{Yu2012}
Yu,~Y.-S.; Wang,~Z.; Zhao,~Y.-P. {Experimental and theoretical investigations
  of evaporation of sessile water droplet on hydrophobic surfaces}. \emph{J.
  Colloid Interface Sci.} \textbf{2012}, \emph{365}, 254--259\relax
\mciteBstWouldAddEndPuncttrue
\mciteSetBstMidEndSepPunct{\mcitedefaultmidpunct}
{\mcitedefaultendpunct}{\mcitedefaultseppunct}\relax
\EndOfBibitem
\bibitem[Hu and Larson(2002)Hu, and Larson]{Hu2002}
Hu,~H.; Larson,~R.~G. {Evaporation of a Sessile Droplet on a Substrate}.
  \emph{J. Phys. Chem. B} \textbf{2002}, \emph{106}, 1334--1344\relax
\mciteBstWouldAddEndPuncttrue
\mciteSetBstMidEndSepPunct{\mcitedefaultmidpunct}
{\mcitedefaultendpunct}{\mcitedefaultseppunct}\relax
\EndOfBibitem
\bibitem[Picknett and Bexon(1977)Picknett, and Bexon]{Picknett1977}
Picknett,~R.; Bexon,~R. {The evaporation of sessile or pendant drops in still
  air}. \emph{J. Colloid Interface Sci.} \textbf{1977}, \emph{61},
  336--350\relax
\mciteBstWouldAddEndPuncttrue
\mciteSetBstMidEndSepPunct{\mcitedefaultmidpunct}
{\mcitedefaultendpunct}{\mcitedefaultseppunct}\relax
\EndOfBibitem
\bibitem[Erbil \latin{et~al.}(2002)Erbil, McHale, and Newton]{Erbil2002}
Erbil,~H.~Y.; McHale,~G.; Newton,~M.~I. {Drop Evaporation on Solid Surfaces:
  Constant Contact Angle Mode}. \emph{Langmuir} \textbf{2002}, \emph{18},
  2636--2641\relax
\mciteBstWouldAddEndPuncttrue
\mciteSetBstMidEndSepPunct{\mcitedefaultmidpunct}
{\mcitedefaultendpunct}{\mcitedefaultseppunct}\relax
\EndOfBibitem
\bibitem[Hughes \latin{et~al.}(2015)Hughes, Thiele, and Archer]{Hughes2015}
Hughes,~A.~P.; Thiele,~U.; Archer,~A.~J. {Liquid drops on a surface: Using
  density functional theory to calculate the binding potential and drop
  profiles and comparing with results from mesoscopic modelling}. \emph{J.
  Chem. Phys.} \textbf{2015}, \emph{142}, 074702\relax
\mciteBstWouldAddEndPuncttrue
\mciteSetBstMidEndSepPunct{\mcitedefaultmidpunct}
{\mcitedefaultendpunct}{\mcitedefaultseppunct}\relax
\EndOfBibitem
\bibitem[Man and Doi(2016)Man, and Doi]{Man2016}
Man,~X.; Doi,~M. {Ring to Mountain Transition in Deposition Pattern of Drying
  Droplets}. \emph{Phys. Rev. Lett.} \textbf{2016}, \emph{116}, 066101\relax
\mciteBstWouldAddEndPuncttrue
\mciteSetBstMidEndSepPunct{\mcitedefaultmidpunct}
{\mcitedefaultendpunct}{\mcitedefaultseppunct}\relax
\EndOfBibitem
\bibitem[Frank and Perr{\'{e}}(2012)Frank, and Perr{\'{e}}]{Frank2012}
Frank,~X.; Perr{\'{e}},~P. {Droplet spreading on a porous surface: A lattice
  Boltzmann study}. \emph{Phys. Fluids} \textbf{2012}, \emph{24}, 042101\relax
\mciteBstWouldAddEndPuncttrue
\mciteSetBstMidEndSepPunct{\mcitedefaultmidpunct}
{\mcitedefaultendpunct}{\mcitedefaultseppunct}\relax
\EndOfBibitem
\bibitem[Ledesma-Aguilar \latin{et~al.}(2014)Ledesma-Aguilar, Vella, and
  Yeomans]{Ledesma-Aguilar2014}
Ledesma-Aguilar,~R.; Vella,~D.; Yeomans,~J.~M. {Lattice-Boltzmann simulations
  of droplet evaporation}. \emph{Soft Matter} \textbf{2014}, \emph{10},
  8267--8275\relax
\mciteBstWouldAddEndPuncttrue
\mciteSetBstMidEndSepPunct{\mcitedefaultmidpunct}
{\mcitedefaultendpunct}{\mcitedefaultseppunct}\relax
\EndOfBibitem
\bibitem[Hirvi and Pakkanen(2006)Hirvi, and Pakkanen]{Hirvi2006}
Hirvi,~J.~T.; Pakkanen,~T.~A. {Molecular dynamics simulations of water droplets
  on polymer surfaces}. \emph{J. Chem. Phys.} \textbf{2006}, \emph{125},
  144712\relax
\mciteBstWouldAddEndPuncttrue
\mciteSetBstMidEndSepPunct{\mcitedefaultmidpunct}
{\mcitedefaultendpunct}{\mcitedefaultseppunct}\relax
\EndOfBibitem
\bibitem[Zhang \latin{et~al.}(2014)Zhang, Leroy, and
  M{\"{u}}ller-Plathe]{Zhang2014}
Zhang,~J.; Leroy,~F.; M{\"{u}}ller-Plathe,~F. {Influence of Contact-Line
  Curvature on the Evaporation of Nanodroplets from Solid Substrates}.
  \emph{Phys. Rev. Lett.} \textbf{2014}, \emph{113}, 046101\relax
\mciteBstWouldAddEndPuncttrue
\mciteSetBstMidEndSepPunct{\mcitedefaultmidpunct}
{\mcitedefaultendpunct}{\mcitedefaultseppunct}\relax
\EndOfBibitem
\bibitem[Stauber \latin{et~al.}(2014)Stauber, Wilson, Duffy, and
  Sefiane]{Stauber2014}
Stauber,~J.~M.; Wilson,~S.~K.; Duffy,~B.~R.; Sefiane,~K. {On the lifetimes of
  evaporating droplets}. \emph{J. Fluid Mech.} \textbf{2014}, \emph{744},
  R2\relax
\mciteBstWouldAddEndPuncttrue
\mciteSetBstMidEndSepPunct{\mcitedefaultmidpunct}
{\mcitedefaultendpunct}{\mcitedefaultseppunct}\relax
\EndOfBibitem
\bibitem[Stauber \latin{et~al.}(2015)Stauber, Wilson, Duffy, and
  Sefiane]{Stauber2015a}
Stauber,~J.~M.; Wilson,~S.~K.; Duffy,~B.~R.; Sefiane,~K. {On the lifetimes of
  evaporating droplets with related initial and receding contact angles}.
  \emph{Phys. Fluids} \textbf{2015}, \emph{27}, 122101\relax
\mciteBstWouldAddEndPuncttrue
\mciteSetBstMidEndSepPunct{\mcitedefaultmidpunct}
{\mcitedefaultendpunct}{\mcitedefaultseppunct}\relax
\EndOfBibitem
\bibitem[Schonhorn \latin{et~al.}(1966)Schonhorn, Frisch, and
  Kwei]{Schonhorn1966}
Schonhorn,~H.; Frisch,~H.; Kwei,~T. {Kinetics of Wetting of Surfaces by Polymer
  Melts}. \emph{J. Appl. Phys.} \textbf{1966}, \emph{37}, 4967\relax
\mciteBstWouldAddEndPuncttrue
\mciteSetBstMidEndSepPunct{\mcitedefaultmidpunct}
{\mcitedefaultendpunct}{\mcitedefaultseppunct}\relax
\EndOfBibitem
\bibitem[Newman(1968)]{Newman1968}
Newman,~S. {Kinetics of wetting of surfaces by polymers; capillary flow}.
  \emph{J. Colloid Interface Sci.} \textbf{1968}, \emph{26}, 209--213\relax
\mciteBstWouldAddEndPuncttrue
\mciteSetBstMidEndSepPunct{\mcitedefaultmidpunct}
{\mcitedefaultendpunct}{\mcitedefaultseppunct}\relax
\EndOfBibitem
\bibitem[Blake and Haynes(1969)Blake, and Haynes]{Blake1969}
Blake,~T.; Haynes,~J. {Kinetics of displacement}. \emph{J. Colloid Interface
  Sci.} \textbf{1969}, \emph{30}, 421--423\relax
\mciteBstWouldAddEndPuncttrue
\mciteSetBstMidEndSepPunct{\mcitedefaultmidpunct}
{\mcitedefaultendpunct}{\mcitedefaultseppunct}\relax
\EndOfBibitem
\bibitem[de~Gennes \latin{et~al.}(2004)de~Gennes, Brochard-Wyart, and
  Qu{\'{e}}r{\'{e}}]{DeGennes2004}
de~Gennes,~P.-G.; Brochard-Wyart,~F.; Qu{\'{e}}r{\'{e}},~D. \emph{{Capillarity
  and Wetting Phenomena}}; Springer: New York, 2004\relax
\mciteBstWouldAddEndPuncttrue
\mciteSetBstMidEndSepPunct{\mcitedefaultmidpunct}
{\mcitedefaultendpunct}{\mcitedefaultseppunct}\relax
\EndOfBibitem
\bibitem[Extrand and Kumagai(1996)Extrand, and Kumagai]{Extrand1996}
Extrand,~C.; Kumagai,~Y. {Contact Angles and Hysteresis on Soft Surfaces}.
  \emph{J. Colloid Interface Sci.} \textbf{1996}, \emph{184}, 191--200\relax
\mciteBstWouldAddEndPuncttrue
\mciteSetBstMidEndSepPunct{\mcitedefaultmidpunct}
{\mcitedefaultendpunct}{\mcitedefaultseppunct}\relax
\EndOfBibitem
\bibitem[Whyman \latin{et~al.}(2008)Whyman, Bormashenko, and Stein]{Whyman2008}
Whyman,~G.; Bormashenko,~E.; Stein,~T. {The rigorous derivation of Young,
  Cassie–Baxter and Wenzel equations and the analysis of the contact angle
  hysteresis phenomenon}. \emph{Chem. Phys. Lett.} \textbf{2008}, \emph{450},
  355--359\relax
\mciteBstWouldAddEndPuncttrue
\mciteSetBstMidEndSepPunct{\mcitedefaultmidpunct}
{\mcitedefaultendpunct}{\mcitedefaultseppunct}\relax
\EndOfBibitem
\bibitem[Young(1805)]{Young1805}
Young,~T. {An Essay on the Cohesion of Fluids}. \emph{Philos. Trans. R. Soc.
  London} \textbf{1805}, \emph{95}, 65--87\relax
\mciteBstWouldAddEndPuncttrue
\mciteSetBstMidEndSepPunct{\mcitedefaultmidpunct}
{\mcitedefaultendpunct}{\mcitedefaultseppunct}\relax
\EndOfBibitem
\bibitem[de~Gennes(1985)]{DeGennes1985}
de~Gennes,~P.~G. {Wetting: statics and dynamics}. \emph{Rev. Mod. Phys.}
  \textbf{1985}, \emph{57}, 827--863\relax
\mciteBstWouldAddEndPuncttrue
\mciteSetBstMidEndSepPunct{\mcitedefaultmidpunct}
{\mcitedefaultendpunct}{\mcitedefaultseppunct}\relax
\EndOfBibitem
\bibitem[Swift and Hohenberg(1977)Swift, and Hohenberg]{Swift1977}
Swift,~J.; Hohenberg,~P.~C. {Hydrodynamic fluctuations at the convective
  instability}. \emph{Phys. Rev. A} \textbf{1977}, \emph{15}, 319--328\relax
\mciteBstWouldAddEndPuncttrue
\mciteSetBstMidEndSepPunct{\mcitedefaultmidpunct}
{\mcitedefaultendpunct}{\mcitedefaultseppunct}\relax
\EndOfBibitem
\bibitem[Goldenfeld(1992)]{Goldenfeld1992}
Goldenfeld,~N.~D. \emph{{Lectures On Phase Transitions And The Renormalization
  Group}}; Perseus Books: New York, 1992; p 209\relax
\mciteBstWouldAddEndPuncttrue
\mciteSetBstMidEndSepPunct{\mcitedefaultmidpunct}
{\mcitedefaultendpunct}{\mcitedefaultseppunct}\relax
\EndOfBibitem
\bibitem[Chaikin and Lubensky(1995)Chaikin, and Lubensky]{Chaikin1995}
Chaikin,~P.~M.; Lubensky,~T.~C. \emph{{Principles of condensed matter
  physics}}; Cambridge University Press, 1995; pp 466--468\relax
\mciteBstWouldAddEndPuncttrue
\mciteSetBstMidEndSepPunct{\mcitedefaultmidpunct}
{\mcitedefaultendpunct}{\mcitedefaultseppunct}\relax
\EndOfBibitem
\bibitem[H{\"{a}}rth and Schubert(2012)H{\"{a}}rth, and Schubert]{Harth2012}
H{\"{a}}rth,~M.; Schubert,~D.~W. {Simple Approach for Spreading Dynamics of
  Polymeric Fluids}. \emph{Macromol. Chem. Phys.} \textbf{2012}, \emph{213},
  654--665\relax
\mciteBstWouldAddEndPuncttrue
\mciteSetBstMidEndSepPunct{\mcitedefaultmidpunct}
{\mcitedefaultendpunct}{\mcitedefaultseppunct}\relax
\EndOfBibitem
\bibitem[Weirich \latin{et~al.}(2017)Weirich, Banerjee, Dasbiswas, Witten,
  Vaikuntanathan, and Gardel]{Weirich2017}
Weirich,~K.~L.; Banerjee,~S.; Dasbiswas,~K.; Witten,~T.~A.; Vaikuntanathan,~S.;
  Gardel,~M.~L. {Liquid behavior of cross-linked actin bundles}. \emph{Proc.
  Natl. Acad. Sci.} \textbf{2017}, \emph{114}, 2131--2136\relax
\mciteBstWouldAddEndPuncttrue
\mciteSetBstMidEndSepPunct{\mcitedefaultmidpunct}
{\mcitedefaultendpunct}{\mcitedefaultseppunct}\relax
\EndOfBibitem
\bibitem[Bonn \latin{et~al.}(2009)Bonn, Eggers, Indekeu, Meunier, and
  Rolley]{Bonn2009}
Bonn,~D.; Eggers,~J.; Indekeu,~J.; Meunier,~J.; Rolley,~E. {Wetting and
  spreading}. \emph{Rev. Mod. Phys.} \textbf{2009}, \emph{81}, 739--805\relax
\mciteBstWouldAddEndPuncttrue
\mciteSetBstMidEndSepPunct{\mcitedefaultmidpunct}
{\mcitedefaultendpunct}{\mcitedefaultseppunct}\relax
\EndOfBibitem
\bibitem[Andrieu \latin{et~al.}(2002)Andrieu, Beysens, Nikolayev, and
  Pomeau]{Andrieu2002}
Andrieu,~C.; Beysens,~D.~A.; Nikolayev,~V.~S.; Pomeau,~Y. {Coalescence of
  sessile drops}. \emph{J. Fluid Mech.} \textbf{2002}, \emph{453},
  427--438\relax
\mciteBstWouldAddEndPuncttrue
\mciteSetBstMidEndSepPunct{\mcitedefaultmidpunct}
{\mcitedefaultendpunct}{\mcitedefaultseppunct}\relax
\EndOfBibitem
\bibitem[Cherry and Holmes(1969)Cherry, and Holmes]{Cherry1969}
Cherry,~B.; Holmes,~C. {Kinetics of wetting of surfaces by polymers}. \emph{J.
  Colloid Interface Sci.} \textbf{1969}, \emph{29}, 174--176\relax
\mciteBstWouldAddEndPuncttrue
\mciteSetBstMidEndSepPunct{\mcitedefaultmidpunct}
{\mcitedefaultendpunct}{\mcitedefaultseppunct}\relax
\EndOfBibitem
\bibitem[{Van Oene} \latin{et~al.}(1969){Van Oene}, Chang, and
  Newman]{VanOene1969}
{Van Oene},~H.; Chang,~Y.~F.; Newman,~S. {The Rheology of Wetting By Polymer
  Melts}. \emph{J. Adhes.} \textbf{1969}, \emph{1}, 54--68\relax
\mciteBstWouldAddEndPuncttrue
\mciteSetBstMidEndSepPunct{\mcitedefaultmidpunct}
{\mcitedefaultendpunct}{\mcitedefaultseppunct}\relax
\EndOfBibitem
\bibitem[Varma \latin{et~al.}(1974)Varma, Sharma, and Varma]{Varma1974}
Varma,~T.; Sharma,~L.; Varma,~S. {Kinetics of spreading of a spherical droplet
  on a smooth surface}. \emph{Surf. Sci.} \textbf{1974}, \emph{45},
  205--212\relax
\mciteBstWouldAddEndPuncttrue
\mciteSetBstMidEndSepPunct{\mcitedefaultmidpunct}
{\mcitedefaultendpunct}{\mcitedefaultseppunct}\relax
\EndOfBibitem
\bibitem[de~Ruijter \latin{et~al.}(1997)de~Ruijter, {De Coninck}, Blake,
  Clarke, and Rankin]{DeRuijter1997}
de~Ruijter,~M.~J.; {De Coninck},~J.; Blake,~T.~D.; Clarke,~A.; Rankin,~A.
  {Contact Angle Relaxation during the Spreading of Partially Wetting Drops}.
  \emph{Langmuir} \textbf{1997}, \emph{13}, 7293--7298\relax
\mciteBstWouldAddEndPuncttrue
\mciteSetBstMidEndSepPunct{\mcitedefaultmidpunct}
{\mcitedefaultendpunct}{\mcitedefaultseppunct}\relax
\EndOfBibitem
\bibitem[Blake(2006)]{Blake2006}
Blake,~T.~D. {The physics of moving wetting lines}. \emph{J. Colloid Interface
  Sci.} \textbf{2006}, \emph{299}, 1--13\relax
\mciteBstWouldAddEndPuncttrue
\mciteSetBstMidEndSepPunct{\mcitedefaultmidpunct}
{\mcitedefaultendpunct}{\mcitedefaultseppunct}\relax
\EndOfBibitem
\bibitem[Seveno \latin{et~al.}(2009)Seveno, Vaillant, Rioboo, Adão, Conti,
  and {De Coninck}]{Seveno2009}
Seveno,~D.; Vaillant,~A.; Rioboo,~R.; Adão,~H.; Conti,~J.; {De Coninck},~J.
  {Dynamics of Wetting Revisited}. \emph{Langmuir} \textbf{2009}, \emph{25},
  13034--13044\relax
\mciteBstWouldAddEndPuncttrue
\mciteSetBstMidEndSepPunct{\mcitedefaultmidpunct}
{\mcitedefaultendpunct}{\mcitedefaultseppunct}\relax
\EndOfBibitem
\bibitem[Blake \latin{et~al.}(1997)Blake, Clarke, {De Coninck}, and
  de~Ruijter]{Blake1997}
Blake,~T.~D.; Clarke,~A.; {De Coninck},~J.; de~Ruijter,~M.~J. {Contact Angle
  Relaxation during Droplet Spreading: Comparison between Molecular Kinetic
  Theory and Molecular Dynamics}. \emph{Langmuir} \textbf{1997}, \emph{13},
  2164--2166\relax
\mciteBstWouldAddEndPuncttrue
\mciteSetBstMidEndSepPunct{\mcitedefaultmidpunct}
{\mcitedefaultendpunct}{\mcitedefaultseppunct}\relax
\EndOfBibitem
\bibitem[Blake and {De Coninck}(2002)Blake, and {De Coninck}]{Blake2002}
Blake,~T.; {De Coninck},~J. {The influence of solid–liquid interactions on
  dynamic wetting}. \emph{Adv. Colloid Interface Sci.} \textbf{2002},
  \emph{96}, 21--36\relax
\mciteBstWouldAddEndPuncttrue
\mciteSetBstMidEndSepPunct{\mcitedefaultmidpunct}
{\mcitedefaultendpunct}{\mcitedefaultseppunct}\relax
\EndOfBibitem
\bibitem[Snoeijer and Andreotti(2013)Snoeijer, and Andreotti]{Snoeijer2013}
Snoeijer,~J.~H.; Andreotti,~B. {Moving Contact Lines: Scales, Regimes, and
  Dynamical Transitions}. \emph{Annu. Rev. Fluid Mech.} \textbf{2013},
  \emph{45}, 269--292\relax
\mciteBstWouldAddEndPuncttrue
\mciteSetBstMidEndSepPunct{\mcitedefaultmidpunct}
{\mcitedefaultendpunct}{\mcitedefaultseppunct}\relax
\EndOfBibitem
\bibitem[Popov(2005)]{Popov2005}
Popov,~Y.~O. {Evaporative deposition patterns: Spatial dimensions of the
  deposit}. \emph{Phys. Rev. E} \textbf{2005}, \emph{71}, 036313\relax
\mciteBstWouldAddEndPuncttrue
\mciteSetBstMidEndSepPunct{\mcitedefaultmidpunct}
{\mcitedefaultendpunct}{\mcitedefaultseppunct}\relax
\EndOfBibitem
\bibitem[Johnson and Dettre(1964)Johnson, and Dettre]{Johnson1964}
Johnson,~R.~E.; Dettre,~R.~H. {Contact Angle Hysteresis. III. Study of an
  Idealized Heterogeneous Surface}. \emph{J. Phys. Chem.} \textbf{1964},
  \emph{68}, 1744--1750\relax
\mciteBstWouldAddEndPuncttrue
\mciteSetBstMidEndSepPunct{\mcitedefaultmidpunct}
{\mcitedefaultendpunct}{\mcitedefaultseppunct}\relax
\EndOfBibitem
\bibitem[Andersen and Br{\'{e}}chet(1996)Andersen, and
  Br{\'{e}}chet]{Andersen1996}
Andersen,~J.~V.; Br{\'{e}}chet,~Y. {Pinning of a solid-liquid-vapor interface
  by stripes of obstacles}. \emph{Phys. Rev. E} \textbf{1996}, \emph{53},
  5006--5010\relax
\mciteBstWouldAddEndPuncttrue
\mciteSetBstMidEndSepPunct{\mcitedefaultmidpunct}
{\mcitedefaultendpunct}{\mcitedefaultseppunct}\relax
\EndOfBibitem
\bibitem[Brandon and Marmur(1996)Brandon, and Marmur]{Brandon1996}
Brandon,~S.; Marmur,~A. {Simulation of Contact Angle Hysteresis on Chemically
  Heterogeneous Surfaces}. \emph{J. Colloid Interface Sci.} \textbf{1996},
  \emph{183}, 351--355\relax
\mciteBstWouldAddEndPuncttrue
\mciteSetBstMidEndSepPunct{\mcitedefaultmidpunct}
{\mcitedefaultendpunct}{\mcitedefaultseppunct}\relax
\EndOfBibitem
\bibitem[Sch{\"{a}}ffer and Wong(1998)Sch{\"{a}}ffer, and Wong]{Schaffer1998}
Sch{\"{a}}ffer,~E.; Wong,~P.-z. {Dynamics of Contact Line Pinning in Capillary
  Rise and Fall}. \emph{Phys. Rev. Lett.} \textbf{1998}, \emph{80},
  3069--3072\relax
\mciteBstWouldAddEndPuncttrue
\mciteSetBstMidEndSepPunct{\mcitedefaultmidpunct}
{\mcitedefaultendpunct}{\mcitedefaultseppunct}\relax
\EndOfBibitem
\bibitem[Ondar{\c{c}}uhu and Piednoir(2005)Ondar{\c{c}}uhu, and
  Piednoir]{Ondarcuhu2005}
Ondar{\c{c}}uhu,~T.; Piednoir,~A. {Pinning of a Contact Line on Nanometric
  Steps during the Dewetting of a Terraced Substrate}. \emph{Nano Lett.}
  \textbf{2005}, \emph{5}, 1744--1750\relax
\mciteBstWouldAddEndPuncttrue
\mciteSetBstMidEndSepPunct{\mcitedefaultmidpunct}
{\mcitedefaultendpunct}{\mcitedefaultseppunct}\relax
\EndOfBibitem
\bibitem[Debuisson \latin{et~al.}(2016)Debuisson, Merlen, Senez, and
  Arscott]{Debuisson2016}
Debuisson,~D.; Merlen,~A.; Senez,~V.; Arscott,~S. {Stick–Jump (SJ)
  Evaporation of Strongly Pinned Nanoliter Volume Sessile Water Droplets on
  Quick Drying, Micropatterned Surfaces}. \emph{Langmuir} \textbf{2016},
  \emph{32}, 2679--2686\relax
\mciteBstWouldAddEndPuncttrue
\mciteSetBstMidEndSepPunct{\mcitedefaultmidpunct}
{\mcitedefaultendpunct}{\mcitedefaultseppunct}\relax
\EndOfBibitem
\bibitem[{De Coninck} \latin{et~al.}(2017){De Coninck}, {Fern{\'{a}}ndez
  Toledano}, Dunlop, and Huillet]{DeConinck2017}
{De Coninck},~J.; {Fern{\'{a}}ndez Toledano},~J.~C.; Dunlop,~F.; Huillet,~T.
  {Pinning of a drop by a junction on an incline}. \emph{Phys. Rev. E}
  \textbf{2017}, \emph{96}, 042804\relax
\mciteBstWouldAddEndPuncttrue
\mciteSetBstMidEndSepPunct{\mcitedefaultmidpunct}
{\mcitedefaultendpunct}{\mcitedefaultseppunct}\relax
\EndOfBibitem
\end{mcitethebibliography}

\end{document}